\documentclass[aps,prl,amsmath,amssymb,reprint, superscriptaddress,showpacs]{revtex4-2}
\usepackage{color, graphicx}     
\usepackage{dcolumn}     
\usepackage{bm}                  
\usepackage[colorlinks=true,urlcolor=blue,linkcolor=blue,citecolor=blue]{hyperref}%
\usepackage[figure,figure*]{hypcap}
\usepackage{amssymb}
\usepackage{latexsym}
\usepackage{amsfonts}
\usepackage{amsmath}
\usepackage{multirow}

\usepackage[dvipsnames]{xcolor}
\usepackage{times}
\usepackage{amsmath}
\UseRawInputEncoding

\begin{document}
	\preprint{APS/123-QED}	
	\makeatletter
	\newcommand*\bigcdot{\mathpalette\bigcdot@{.3}}
	\newcommand*\bigcdot@[2]{\mathbin{\vcenter{\hbox{\scalebox{#2}{$\m@th#1\bullet$}}}}}
	\makeatother
		
	\title{Aperiodic-quasiperiodic-periodic properties and topological
		transitions  in  twisted  nested Moir\texorpdfstring{$\mathbf{\acute{e}}$}{e} patterns}

	\author{Peng Peng}
    \affiliation{Key~Laboratory~for~Micro/Nano-Optoelectronic~Devices~of~Ministry~of~Education,~School~of~Physics~and~Electronics, Hunan~University,~Changsha~410082,~China} 
        \author{Yuchen Peng}
        \affiliation{Key~Laboratory~for~Micro/Nano-Optoelectronic~Devices~of~Ministry~of~Education,~School~of~Physics~and~Electronics, Hunan~University,~Changsha~410082,~China} 
       \author{Aoqian Shi}
       \affiliation{Key~Laboratory~for~Micro/Nano-Optoelectronic~Devices~of~Ministry~of~Education,~School~of~Physics~and~Electronics, Hunan~University,~Changsha~410082,~China} 
       \author{Xiaogen Yi}
\affiliation{State~Key~Laboratory~for~Mesoscopic~Physics~and~Frontiers~Science ~Center~for~Nano-optoelectronics,~School~of Physics,~Peking~University,~Beijing~100000,~China}%
       \author{Yizhou Wei}
       \affiliation{Key~Laboratory~for~Micro/Nano-Optoelectronic~Devices~of~Ministry~of~Education,~School~of~Physics~and~Electronics, Hunan~University,~Changsha~410082,~China} 
       \author{Jianjun Liu}
    \email{Corresponding author: jianjun.liu@hnu.edu.cn}
       \affiliation{Key~Laboratory~for~Micro/Nano-Optoelectronic~Devices~of~Ministry~of~Education,~School~of~Physics~and~Electronics, Hunan~University,~Changsha~410082,~China} 
       \affiliation{Greater~Bay~Area~Institute~for~Innovation,~Hunan~University, ~Guangzhou~511300,~China}

	\date{\today}
	
	\begin{abstract}
	The Moir$\mathrm{\acute{e}}$ patterns generated by altering the structural parameters in a two or more layers of periodic materials, including single-layer structure, interlayer stacking, and twisting parameters, exhibit prosperous topological physical properties. However, the intricate characteristics of twisted  nested Moir$\mathrm{\acute{e}}$  patterns and their relationship with topological transitions remain unclear. In this Letter, based on the proposed twisted nested photonic crystal (TNPC), we derive its spatial geometric functions (SGFs), aperiodic-quasiperiodic-periodic  properties in twisted  nested Moir\texorpdfstring{$\mathbf{\acute{e}}$}{e} patterns, and the SSH$\varphi$ Hamiltonian. We reveal the intrinsic correlation between twisted  nested Moir$\mathrm{\acute{e}}$  patterns and topological transitions, obtaining higher-order topological states (HOTSs) with $C_{2z}$ symmetry. This work will provide theoretical references for the design and application of twisted topological PC and their devices. 
	\end{abstract}
	\maketitle
    Twisted photonic crystals (TPCs) with optical flat bands, which serve as a photonic analogue platform to explore the electronic properties of twisted graphene~\cite{RN1}, have attracted considerable attention and have been empirically proven to manifest a diverse array of photonic and polarization phenomena, such as bound states in the continuum~\cite{RN2,RN3},  waveguides \cite{RN4}, encoding and decoding of complex information~\cite{RN5}, localization-delocalization transition~\cite{RN6}, lasers~\cite{RN7,RN8}, enhanced nonlinear optical effects~\cite{RN9}, slow light~\cite{RN10,RN11}, Moir$\mathrm{\acute{e}}$ patterns of circular birefringence~\cite{RN12}. Recently, relevant experimental and theoretical researches have investigated edge states or high-order topological states (HOTSs) based on TPCs~\cite{RN4,RN13,RN14,RN15}, which indicates that TPCs hold promise as a new platform in topological physics for achieving customized exotic light-matter interactions. Moir$\mathrm{\acute{e}}$ patterns, generated by stacking and twisting two or more layers of periodic materials, introduce additional lattice distortions, which leading to the formation of a new lattice potential and inducing peculiar topological properties. Therefore, Moir$\mathrm{\acute{e}}$ patterns play a pivotal role in researches of TPCs~\cite{RN1,RN16}, can be generated by varying single-layer structure, interlayer stacking, and twisting parameters. Such Moir$\mathrm{\acute{e}}$ patterns are subject to various controls, including lattice translations~\cite{RN16,RN17,RN18}, lattice constants~\cite{RN19} and twist angles~\cite{RN13,RN20}. By modulating Moir$\mathrm{\acute{e}}$ patterns, precise control over the topological states in PCs can be achieved, providing a deeper understanding and utilization of topological properties. 
	
    Currently, TPCs provide new opportunities for understanding and utilizing topological properties in the field of optics  ~\cite{RN1,RN7,RN20,RN21,RN22}. However, to date, the correlation between the structural parameters of TPCs and the resulting properties of Moir$\mathrm{\acute{e}}$ patterns, as well as their connection to topological physics, remains unexplored. Therefore, establishing specific spatial geometric functions (SGFs) for TPCs and linking them to topological physics will provide unique insights into understanding topological optical phenomena. On the other hand, research related to double-layer lattices has mainly focused on structures formed by stacking two honeycomb lattices or stacking honeycomb lattices with other lattice structures, creating approximately periodic structures~\cite{RN7,RN12,RN14,RN23}. In contrast, PCs, as artificial microstructures, exhibit a richer variety of lattice types, lattice constants, scatterer sizes, shapes, and arrangements~\cite{RN24,RN25,RN26,RN27}, resulting in more exotic Moir$\mathrm{\acute{e}}$ patterns. Therefore, the introduction of novel structures is bound to enrich Moir$\mathrm{\acute{e}}$ pattern researches. Equally important is that Moir$\mathrm{\acute{e}}$ patterns in PCs are controlled by the twist angle. However, it is challenging to change the twist angle between bilayer PCs in finished devices, as twisting involves mechanical motion, making it inconvenient, difficult to achieve, and unstable, especially for small and precise angle twists~\cite{RN28}. Therefore, exploring alternative methods to directly control the lattice potential and thereby influence topological physical holds significant potential for practical applications. 
		
        In this Letter, a TNPC was constructed based on the stacked-twisted single-layer nested honeycomb lattices, from which its SGFs were derived. The aperiodic-quasiperiodic-periodic properties  in twisted  nested Moir\texorpdfstring{$\mathbf{\acute{e}}$}{e} patterns were elucidated by solving commensurate condition. The SSH$\varphi$ Hamiltonian and the relationship between couplings and topological states were obtained through the introduction of multi-degree-of-freedom coupling such as lattice-nesting-stacking-twisting. Topological transitions were achieved by tuning the nested factor rather than the twist angle. Finally, HOTSs were realized through the arrangement in a box-shaped structure.
	
\begin{figure} 
		\centerline{\includegraphics[scale=.12]{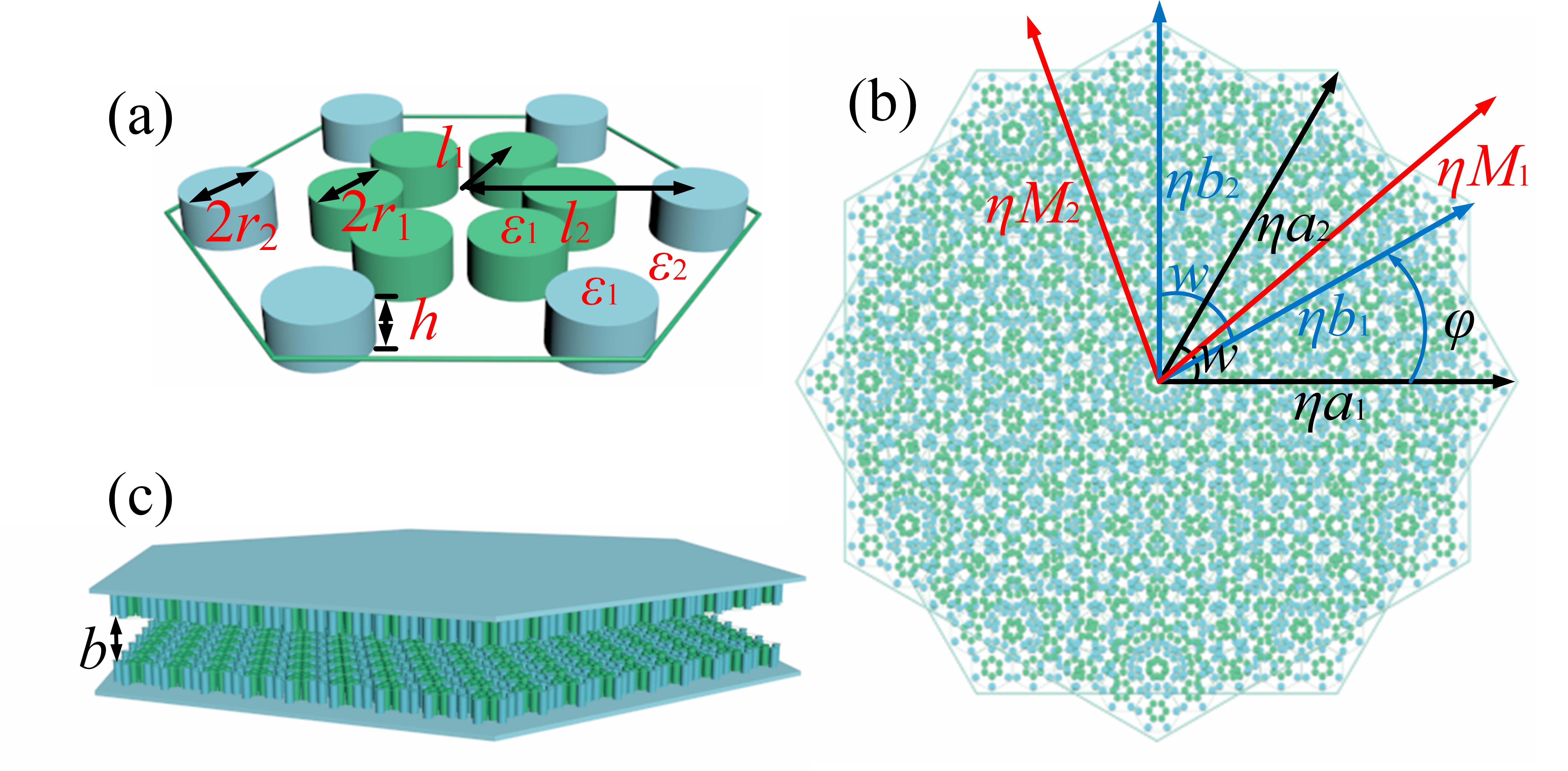}}
		\caption{\label{Fig.1} (a) The unit cell of the single-layer nested honeycomb lattice PC. The inner and outer rings of scatterers have radii denoted as $r_1$  and $r_2$, and distances from the cell center labeled as $l_1$ and $l_2$. The scatterer height is represented by $h$, the lattice constant is denoted as $a$, and the scatterer material is InGaAsP ($\epsilon_1=$10.89+0.01i), while the surrounding material is air $\epsilon_2=$1. (b) Model of the TNPC with a twist angle $\varphi$. The lattice vectors for the lower-layer, upper-layer, and TNPC are denoted as  $\bm{a}_1$,   $\bm{a}_2$,  $\bm{b}_1$,  $\bm{b}_2$,  $\bm{M}_1$ and  $\bm{M}_2$, where $w$ is the angle between two vectors in the same layer, and $\eta$η represents the number of unit cells included along the edge of the honeycomb lattice. (c) Interlayer distance $b$.}
	\end{figure}

        The constructed TNPC is depicted in Fig.~\ref{Fig.1}. The structure of the TNPC is obtained by periodically tiling the unit cell showned in Fig.~\ref{Fig.1}(a), and is divided into two layers, as illustrated in Fig.~\ref{Fig.1}(c). During the twist operation, the lower-layer remains fixed, and only the upper-layer of the TNPC undergoes twist. Referring to Fig.~\ref{Fig.1}(b), the Wood notations~\cite{RN29} is introduced to represent the bilayer Moir$\mathrm{\acute{e}}$ lattice: $\mathrm{i}\left[\left(\boldsymbol{b}_1 /\boldsymbol{a}_1\right) \times\left(\boldsymbol{b}_2 / \boldsymbol{a}_2\right)\right] R_{\varphi}$. Due to the complexity and nonlinearity of the layered structure, a composite function is needed to describe the SGFs. Because of the periodicity, the SGFs of a single-layer can be expressed in the form of a complex Fourier series using the infinite Fourier series method,
        \begin{equation}
		\label{eq.1}
		f_{1(\boldsymbol{r})}=\sum_{m_1, m_2} C_{m_1, m_2}^1 \exp \left[i\left(\frac{2 \pi}{\boldsymbol{a}_1} m_1+\frac{2 \pi}{\boldsymbol{a}_2} m_2\right) \boldsymbol{r}\right.
	\end{equation}
     \quad{The orthogonality of direct and reciprocal lattice vectors manifests as a Dirac $\delta$ function with a periodicity of 2$\mathrm\pi$, denoted as $\boldsymbol{a} \cdot \boldsymbol{G}=2\pi n$, where $\bm{G}$ represents the reciprocal lattice vector, and $n$ is an arbitrary integer, therefore,}
         \begin{equation}
		\label{eq.2}
		f_{1(\boldsymbol{r})}=\sum_{m_1, m_2} C_{m_1, m_2}^1 \exp \left[i\left(m_1 \boldsymbol{G}_{\boldsymbol{a}_1}+m_2 \boldsymbol{G}_{\boldsymbol{a}_2}\right)\right]\boldsymbol{r}
	\end{equation}
     \quad Analogously, the SGFs for the upper-layer can be expressed as:
         \begin{equation}
		\label{eq.3}
		f_{2(\boldsymbol{r})}=\sum_{m_1, m_2} C_{m_1, m_2}^2 \exp \left[i\left(m_1 \boldsymbol{G}_{\boldsymbol{b}_1}+m_2 \boldsymbol{G}_{\boldsymbol{b}_2}\right)\right]\boldsymbol{r}
	\end{equation}
    here, $\ C_{m_1, m_2}^1$ and $\ C_{m_1, m_2}^2$ represents a complex coefficient, and $m_1$ as well as $m_2$ are both integers. Consequently, employing convolution methods $
    F(\boldsymbol{r})=\sum_{\xi=-\infty}^{+\infty} f(\xi) f(\boldsymbol{r}-\xi)
    $ enables the derivation of the SGFs in the TNPC as,
       \begin{equation}\label{eq.4}
        \hspace{-4mm}
     \resizebox{.95\hsize}{!}{
       $\begin{aligned}
    F_{M(\boldsymbol{r})} & =f_{1(\boldsymbol{r})} * f_{2(\boldsymbol{r})} \\
    & =\sum_{\xi} \sum_{m_1, m_2} C_{m_1, m_2}^1 \exp \left[i\left(m_1 G_{a_1}+m_2 G_{a_2}\right)\right] \xi \\&~\cdot \sum_{m_1 m_2} C_{m_1, m_2}^1 \exp\left[i\left(m_1\boldsymbol{ G}_{b_1}+m_2 \boldsymbol{G}_{b_2}\right)\right](\boldsymbol{r}-\xi) \\
    & =a_{m_1, m_2} \exp \left[i\left(m_1 \boldsymbol{G}_{M1} + m_2\boldsymbol{G}_{M2}\right)\right ]\boldsymbol{r}
    \end{aligned}$
    }
    \end{equation}
    where, $\boldsymbol{G}_{M1}=\boldsymbol{G}_{b1}-\boldsymbol{G}_{a1}$,$\boldsymbol{G}_{M2}=\boldsymbol{G}_{b2}-\boldsymbol{G}_{a2}$ represent the reciprocal lattice vector of the TNPC, and in the equation,

    \begin{equation}\label{eq.5}
    	\hspace{-4mm}
    	\resizebox{.95\hsize}{!}{
    		$\begin{aligned}
    			a_{m_1, m_2} = \sum_{\xi} &\sum_{m_1} \sum_{m_2} C_{m_1, m_2}^1 C_{m_1, m_2}^2 \exp\left\{ -i\left[(m_1 \boldsymbol{G}_{M1} + m_2 \boldsymbol{G}_{M2})\right]\xi \right\} \\
    			&\cdot \exp\left[ i\left(m_1 \boldsymbol{G}_{a1} + m_2 \boldsymbol{G}_{a2}\right)\right] \boldsymbol{r}
    		\end{aligned}$
    	}
    \end{equation}

    Therefore, from Eqs.~\eqref{eq.4} and ~\eqref {eq.5}, it can be deduced that the SGFs of the TNPC are intimately connected to the single-layer structural and the bilayer twisting parameters. In particular, $\ a_{m_1, m_2}$ varies with changes in the single-layer structural and bilayer twisting parameters, while $\ \boldsymbol{G}_{M_1}, \boldsymbol{G}_{M 2}$ characterizes the aperiodic-quasiperiodic-periodic properties.
 
    The commensurate condition of twisted  nested Moir\texorpdfstring{$\mathbf{\acute{e}}$}{e} patterns play a significant role in the fields of stacked periodic pattern and interference researches. Simultaneously, to further explore the variation with the twist, the commensurate condition and the aperiodic-quasiperiodic-periodic properties  are detailed in Fig.~\ref{Fig.2}. The dynamic demonstration of the TNPC with varying twist angles is presented in Sec. A in Supplemental Material.
\begin{figure*} 	
   {\includegraphics[width=0.8\linewidth]{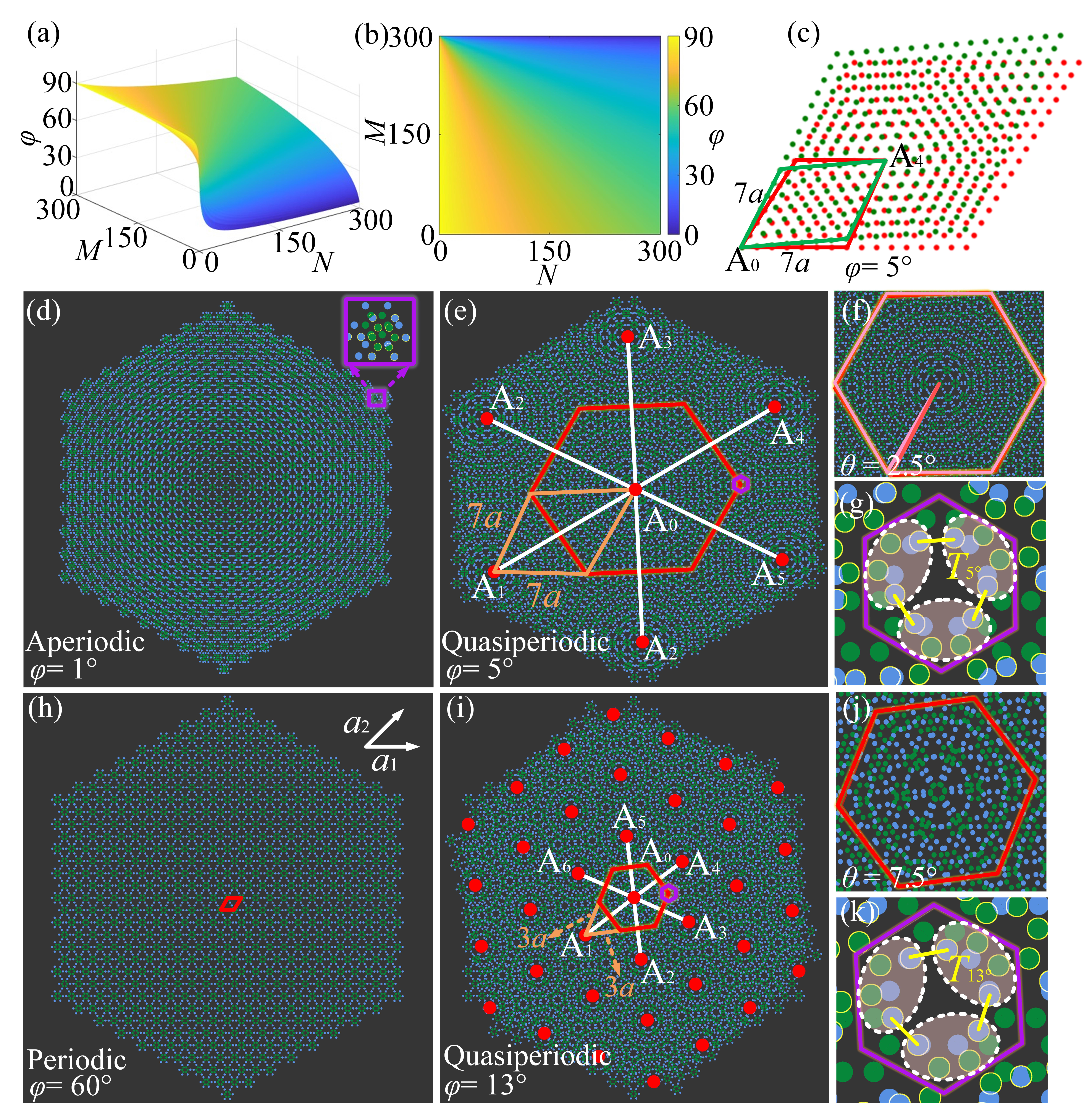}} \\ 
		\caption{\label{Fig.2} Aperiodic-quasiperiodic-periodic properties in twisted  nested Moir\texorpdfstring{$\mathbf{\acute{e}}$}{e} patterns. (a)-(b) Commensurate condition. (c) Schematic representation of twist based on the equivalent substitution method, red and green lattice points indicate the coordinates of the lower-layer (fixed) and upper-layer PC (after twist), respectively. Twisted  nested Moir\texorpdfstring{$\mathrm{\acute{e}}$}{e} patterns under different twist angles and their locally enlarged effects:(d) Aperiodic properties:  $\varphi =1^\circ$, (e)-(g) and (i)-(k) Quasiperiodic properties:  $\varphi =5^\circ$ and  $\varphi =13^\circ$, $T_\varphi$  denotes the nearest-neighbor distance in adjacent stacking distribution regions under different twist. (h) Periodic properties: $\varphi =60^\circ$. }
	\end{figure*}
 
     By employing the twisted transformation of vectors $\bm{a}_1$, $\bm{a}_2$ to $\bm{b}_1$, $\bm{b}_2$ as depicted in Fig.~\ref{Fig.1}(b) and solving the second-order Diophantine equations, the detailed derivation is provided in Sec. B in Supplemental Material. The ultimate commensurate condition is expressed as,
  \begin{equation}
		\label{eq.6}
		\varphi=\frac{1}{2} \cos ^{-1}\left[\frac{-\left(3 M^2+6 M N-3 N^2\right) p_{2(\beta)}}{\left(3 M^2+6 M N+3 N^2\right) p_{1(\beta)}}\right]
	\end{equation}
    Here, \(M\) and \(N\) are integers, and \(\varphi \in [0,2\pi]\). Since the TNPC is formed by stacking and twisting of two identical nested honeycomb lattices, it follows that \(p_{1(\beta)} = p_{2(\beta)}\), \(q_1 = q_2\). As revealed by the commensurate condition shown in Figs.~\ref{Fig.2}(a) and ~\ref{Fig.2}(b), when \(M/N \to 0, \varphi \to 0\) or \(\pi/2\), and when \(M/N \to 1, \varphi \to \pi/3\), correspond to the diagonal part. This indicates that the TNPC exhibits the same twisted  nested Moir\texorpdfstring{$\mathbf{\acute{e}}$}{e}  pattern when \(\varphi' = \varphi + n \pi / 3, (n = 0,1,2,3,4,5)\),
    as demonstrated in Figs.~\ref{Fig.2}(d)-\ref{Fig.2}(k). This behavior is attributed to the $C_6$ symmetry. The modulation of the twist induces prosperous aperiodic-quasiperiodic-periodic properties in twisted  nested Moir\texorpdfstring{$\mathbf{\acute{e}}$}{e}  patterns. However, in general, a higher number of selected unit cells tends to offer a more visually explicit observation of Moir$\mathrm{\acute{e}}$ pattern. Nevertheless, an excessive number of unit cells makes it challenging to distinguish between their aperiodic-quasiperiodic-periodic properties. Therefore, a employing the method of equivalent substitution, each unit cell of the TNPC is regarded as a lattice point in Fig. ~\ref{Fig.2}(c), as evident from the rotation,
 \begin{equation}
		\label{eq.7}
		\left[\begin{array}{l}
A_{\mu x} \\
A_{\mu y}
\end{array}\right]=\left[\begin{array}{ll}
\cos \varphi & \sin \varphi \\
\sin \varphi & \cos \varphi
\end{array}\right]\left[\begin{array}{l}
A_{x_0} \\
A_{y_0}
\end{array}\right]
	\end{equation}
here, $\mu=$0,1,2,3,4,5,6, $\left(A_{\mu x}, A_{\mu y}\right)$ and ~$\left(A_{0 x}, A_{0 y}\right)$ represent the coordinates of the upper-layer of TNPC after twist and the coordinates before twist, respectively. As showned in Fig.~\ref{Fig.2}(c), through the coordinates of each lattice point, one can accurately and intuitively discern its aperiodic-quasiperiodic-periodic properties. The twisted  nested Moir$\mathrm{\acute{e}}$  patterns depicted in Fig.~\ref{Fig.2}(d) exhibit disorder, making it challenging to find stacking distribution regions (SDRs) consistent with the central alignment, particularly in the case of small-angle twists, thus displaying aperiodic properties. The properties of the twisted  nested Moir$\mathrm{\acute{e}}$ patterns in Figs. ~\ref{Fig.2}(c)-\ref{Fig.2}(g) and ~\ref{Fig.2}(i)-\ref{Fig.2}(k) are as follows,
 
$\text { i) }$Rotational symmetry in quasiperiodicity.
Under different twist angles, it exhibits rotational symmetry. Particularly, when $\varphi =30^\circ$, the rotational symmetry is most prominent, exhibiting  $C_{12}$ symmetry, as detailed in Sec. C in Supplemental Material.

$\text { ii) }$Quasi-translation symmetry in quasiperiodicity.
When $\varphi =5^\circ$, twisted  nested Moir$\mathrm{\acute{e}}$  patterns of the TNPC, consisting of (601$\times$2) unit cells, appear with identical SDRs spaced at intervals of $7\sqrt{3}a$
. It is noteworthy that these SDRs are not strictly identical, suggesting that the twisted  nested Moir$\mathrm{\acute{e}}$  pattern resides between the disorder of aperiodicity and the translation symmetry of periodicity. It manifests as twisted  nested Moir$\mathrm{\acute{e}}$  patterns that undergo quasi-translation symmetry, resembling a region enclosed by a regular hexagon rotated by $\theta=\varphi/2 $ on the basis of the original hexagon (as depicted in Fig.~\ref{Fig.2}(f)), undergoing periodic tiling once again. This phenomenon is referred to as "quasi-translation symmetry." Further exploration under large-angle twists involves constructing the TNPC with an equal number of unit cells rotated by $\varphi =13^\circ$, resulting in identical SDRs spaced at intervals of  $3\sqrt{3}a$, as illustrated in Fig. ~\ref{Fig.2}(j), though with fewer geometric pattern features. 

    $\text { iii) }$Long-range oder in quasiperiodicity.
    Connecting the periodically repeated regions with straight lines forms a parallelogram grid, indicating that the twisted  nested Moir$\mathrm{\acute{e}}$  patterns exhibit a long-range ordered arrangement, implying long-range order in quasiperiodicity. In summary, twisted  nested Moir$\mathrm{\acute{e}}$  patterns in the TNPC, composed of an identical number of unit cells, indicate that with an increase in the twist $\varphi < 30^\circ$
    , the spacing between identical SDRs decreases, and the number of included scatterers diminishes. This suggests that, under the same size conditions, more complex geometric pattern features are manifested.

    Additionally, at the edge of SDRs for two different twist angles, there are four scatterers each extracted from six single-layer nested honeycomb lattice unit cells, as shown in Figs. ~\ref{Fig.2}(g) and \ref{Fig.2}(k). $T_{13^{\circ}}>T_{5^{\circ}}$, within one rotation period ($\varphi < 30^\circ$), this indicates that as the twist angle increases, the nearest-neighbor distance $T_\varphi$ becomes larger, offering new insights for Moir$\mathrm{\acute{e}}$ pattern researches. The twisted  nested Moir$\mathrm{\acute{e}}$  patterns with periodic properties are depicted in Fig. ~\ref{Fig.2}(h), exhibiting translational symmetry,  $C_6$   rotational symmetry, and mirror symmetry during  $\varphi^{\prime}=\varphi+n \pi / 3$. 

\begin{figure*}[t]
    {\includegraphics[width=0.8\linewidth]{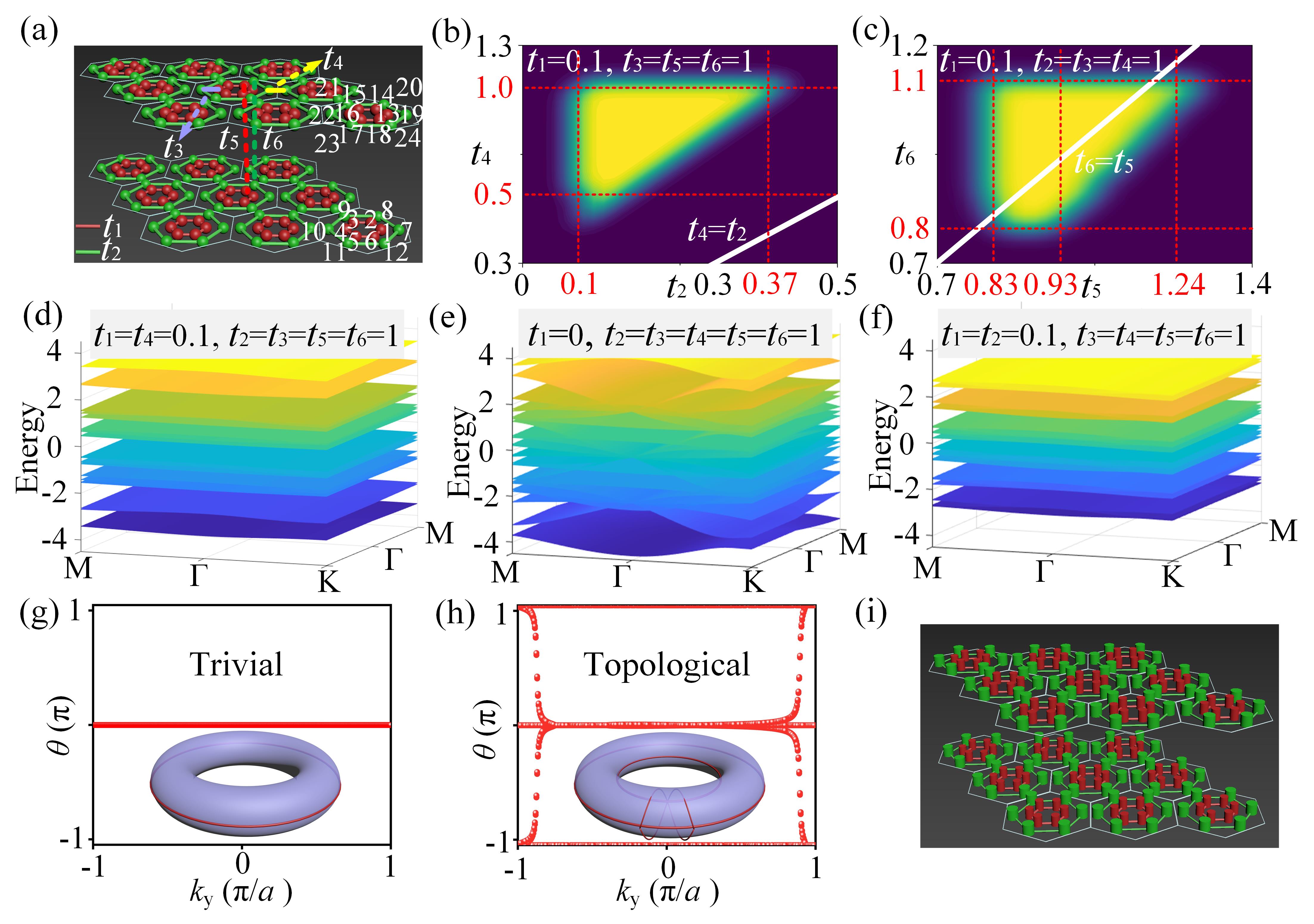}} \\ 
      \caption{\label{Fig.3}(a) The model of the twisted nested honeycomb lattice, where $t_1$, $t_2$, and $t_3$ represent the intra-cell coupling between the inner circles (red spheres), outer circles (green spheres), and inner and outer circles (green and red spheres), respectively. $t_4$ denotes the inter-cell coupling between the nearest-neighbor lattice points, while $t_5$ and $t_6$ represent the inter-layer coupling for two layers of inner circles and two layers of outer circles, respectively. (b) The phase diagrams of $t_2$ and $t_4$ for achieving topological states.(c) The phase diagrams of $t_5$ and $t_6$ for realizing topological states; (d)-(f) The band structures in trivial, degenerate, and topological states. (g)-(h) correspond to the Wilson loops for Figs.~\ref{Fig.3}(d) and ~\ref{Fig.3}(f), with the inset below illustrating that the two-dimensional Brillouin zone is equivalent to a torus under periodic boundary conditions for Bloch states, where the red line represents the winding number of the torus. (i) The TNPC corresponding to the twisted honeycomb lattice.}
\end{figure*}

    In conclusion, different twist angles result in distinct properties of twisted  nested Moir$\mathrm{\acute{e}}$  patterns, including the disorder of aperiodicity, the rotational, quasi-translation symmetry, and long-range order of quasi-periodicity, as well as the translational, rotational, and mirror symmetries of periodicity. Investigating the transition from disorder to various symmetries and ordered patterns in the aperiodic to quasi-periodic to periodic properties provides new insights into understanding the optical behavior of artificial structured materials.
    Through varying the coupling strength, the lattice potential can be further modified, ultimately influencing its topological properties. To investigate the relationship between the TNPC and topological transitions, a tight-binding model with multi-degree-of-freedom coupling such as lattice-nesting-stacking-twisting, is introduced to derive the SSH$\varphi$  Hamiltonian, topological conditions, and band structures, as illustrated in Fig.~\ref{Fig.3}.

    The Hamiltonian of the twisted nested honeycomb lattice can be expressed as,
 \begin{equation}\label{eq.8}
    H=\left[\begin{array}{ll}
    H_m & H_{\varphi} \\
    H_{\varphi} & H_m
\end{array}\right]
\end{equation}
 Where $H_m$ and $H_\varphi$ represent the Hamiltonians for the single-layer nested honeycomb lattice and the interlayer twisted part, respectively. According to Fig. ~\ref{Fig.3}(a), the Hamiltonian for the single-layer nested honeycomb lattice can be described as,
 
  \begin{equation}\label{eq.9}
 	\hspace{-4mm}
 	\resizebox{.95\hsize}{!}{%
 		$\mathit{H}_m = \left[%
 		\begin{array}{cccccccccccc}
 			0 & t_1 & 0 & 0 & 0 & t_1 & t_3 & 0 & 0 & 0 & 0 & 0 \\
 			t_1 & 0 & t_1 & 0 & 0 & 0 & 0 & t_3 & 0 & 0 & 0 & 0 \\
 			0 & t_1 & 0 & t_1 & 0 & 0 & 0 & 0 & t_3 & 0 & 0 & 0 \\
 			0 & 0 & t_1 & 0 & t_1 & 0 & 0 & 0 & 0 & t_3 & 0 & 0 \\
 			0 & 0 & 0 & t_1 & 0 & t_1 & 0 & 0 & 0 & 0 & t_3 & 0 \\
 			t_1 & 0 & 0 & 0 & t_1 & 0 & t_1 & 0 & 0 & 0 & 0 & t_3 \\
 			t_3 & 0 & 0 & 0 & 0 & t_2 & 0 & t_2 & 0 & t_4e^{ik \delta_1} & 0 & t_2 \\
 			0 & t_3 & 0 & 0 & 0 & 0 & t_2 & 0 & t_2 & 0 & t_4e^{ik \delta_3} & 0 \\
 			0 & 0 & t_3 & 0 & 0 & 0 & 0 & t_2 & 0 & t_2 & 0 & t_4e^{ik \delta_2} \\
 			0 & 0 & 0 & t_3 & 0 & 0 & t_4e^{-ik \delta_3} & 0 & t_2 & 0 & t_2 & 0 \\
 			0 & 0 & 0 & 0 & t_3 & 0 & 0 & t_4e^{-ik \delta_3} & 0 & t_2 & 0 & t_2 \\
 			0 & 0 & 0 & 0 & 0 & t_3 & t_2 & 0 & t_4e^{-ik \delta_2} & 0 & t_2 & 0
 		\end{array}%
 		\right]$%
 	}
 \end{equation}
    Where $\delta_1=(1,0)a, \delta_2=(-1/2,\sqrt{3}/2)a,\delta_3=(1/2,\sqrt{3}/2)a$ . It is noteworthy that  only nearest-neighbor coupling is considered here, as the coupling strength rapidly decays with the increasing distance ($t\propto1/l^3$), Consequently, The influence of next-nearest-neighbor coupling is negligible on the dispersion relation, playing a minor role in altering topological properties~\cite{RN30,RN31}. To further obtain the interlayer twisted part of the Hamiltonian, the distorted three direction vectors need to be obtained. The vector relationship between lattice points between the two layers is described as: $\delta_{\varphi, i}=\left[T_d * R_{\varphi}\right] \delta_{i=1,2,3}$, where $T_d$ denotes the translation transformation matrix, and $R_\varphi$ represents the rotation transformation matrix. Therefore, the three direction vectors are characterized by，   
\begin{small} 
\begin{equation} \label{eq.10}
     \delta_{\varphi, 1}=(\cos \varphi-1, \sin \varphi) \boldsymbol{a}
\end{equation}
\end{small}
\begin{small} 
\begin{equation}\label{eq.11}
    \hspace{-2mm}
    \boldsymbol{\delta}_{\varphi,2}=[(-\cos \varphi-1)-\sqrt{3} \sin \varphi, -\sin \varphi+\sqrt{3} (
    \cos\varphi-1)] \boldsymbol{a} / 2
\end{equation}
\end{small}
\begin{small}
\begin{equation}\label{eq.12}
   \hspace{-2mm}
    \boldsymbol{\delta}_{\varphi,3}=[(\cos \varphi-1)-\sqrt{3} \sin \varphi, \sin \varphi+\sqrt{3} (
    \cos\varphi-1)] \boldsymbol{a} / 2
\end{equation}
\end{small}
	
	Substituting Eqs. ~\eqref{eq.11}-\eqref{eq.12} into the interlayer twist Hamiltonian,
       \begin{equation}\label{eq.13}
  \begin{aligned}
    H_{\varphi} = &\operatorname{diag}([ t_5 e^{i k \delta \varphi,1},t_5 e^{i k \delta \varphi,3 },t_5 e^{i k \delta \varphi,2},t_5 e^{i k \delta \varphi,1},\\
    &t_5 e^{i k \delta \varphi,3},t_5 e^{i k \delta \varphi,2},t_6 e^{i k \delta \varphi,1},t_6 e^{i k \delta \varphi,3},\\
    &t_6 e^{i k \delta \varphi,2},t_6 e^{i k \delta \varphi,1}, t_6 e^{i k \delta \varphi,3},t_6 e^{i k \delta \varphi,2}])
  \end{aligned}
\end{equation}
    $H_{\varphi} \in \mathbb{C}^{12 \times 12}, H \in \mathbb{C}^{24 \times 24}$,$\mathbb{C}$ represents the set of complex numbers. The Hamiltonian matrix for the twisted nested honeycomb lattice is a $24\times24$ matrix, corresponding to 24 bands. Each set of 12 bands exhibits symmetry with respect to $E=0$, as illustrated in Figs. ~\ref{Fig.3}(d)-~\ref{Fig.3}(f). Analyzing the Wilson loops~\cite{RN32,RN33} in Figs. ~\ref{Fig.3}(d) and ~\ref{Fig.3}(f), the winding numbers are determined to be 0 and 2, respectively. Therefore, Fig. ~\ref{Fig.3}(g) corresponds to trivial states, while Fig. ~\ref{Fig.3}(h) corresponds to topological states. 

For comprehensively understanding of the correlation between single-layer structural parameters, bilayer twist parameters, and coupling strength with topological states, a matrix tuning algorithm was employed to solve their relationship, as depicted in Figs. ~\ref{Fig.3}(b)-\ref{Fig.3}(c). Under the conditions $t_1=0.1, t_3=t_5=t_6=1$, the system exhibits topological states when $t_5 \in[0.83,1.24], t_6 \in[0.8,1.1]$ in the yellow-highlighted region of Fig. ~\ref{Fig.3}(b) ($t_4>t_2$), indicating inter-cell coupling over intra-cell coupling. Similarly, under the conditions $t_1=0.1, t_2=t_3=t_4=1$, topological states are observed in the highlighted yellow region of Fig.~\ref{Fig.3}(c) when  $t_2 \in[0.1,0.37], t_4 \in[0.5,1]$  . In this case, two scenarios need to be considered: ⅰ) When the yellow-highlighted region is on the right side of the function $t_6=t_5$ (indicating $t_6<t_5$), it means that after twist, the distance between the outer rings of the upper and lower layers is greater than that between the inner rings. Conversely, when the yellow-highlighted region is on the left side of the function $t_6=t_5$ (indicating $t_6>t_5$), it implies that after twist, the distance between the outer rings of the upper and lower layers is smaller than that between the inner rings. However, generally, the distance between the outer rings of the upper and lower layers is not expected to be smaller than that between the inner rings. The coupling depends not only on the distance between two lattice sites but also on their potential energy. In TNPCs, different-sized scatterers can be used for nesting and twist, allowing for the adjustment of potential energy and lattice distances to better construct the SSH$\varphi$  Hamiltonian.
 
\begin{figure}[!t]\centering
   \includegraphics[width=1\linewidth]{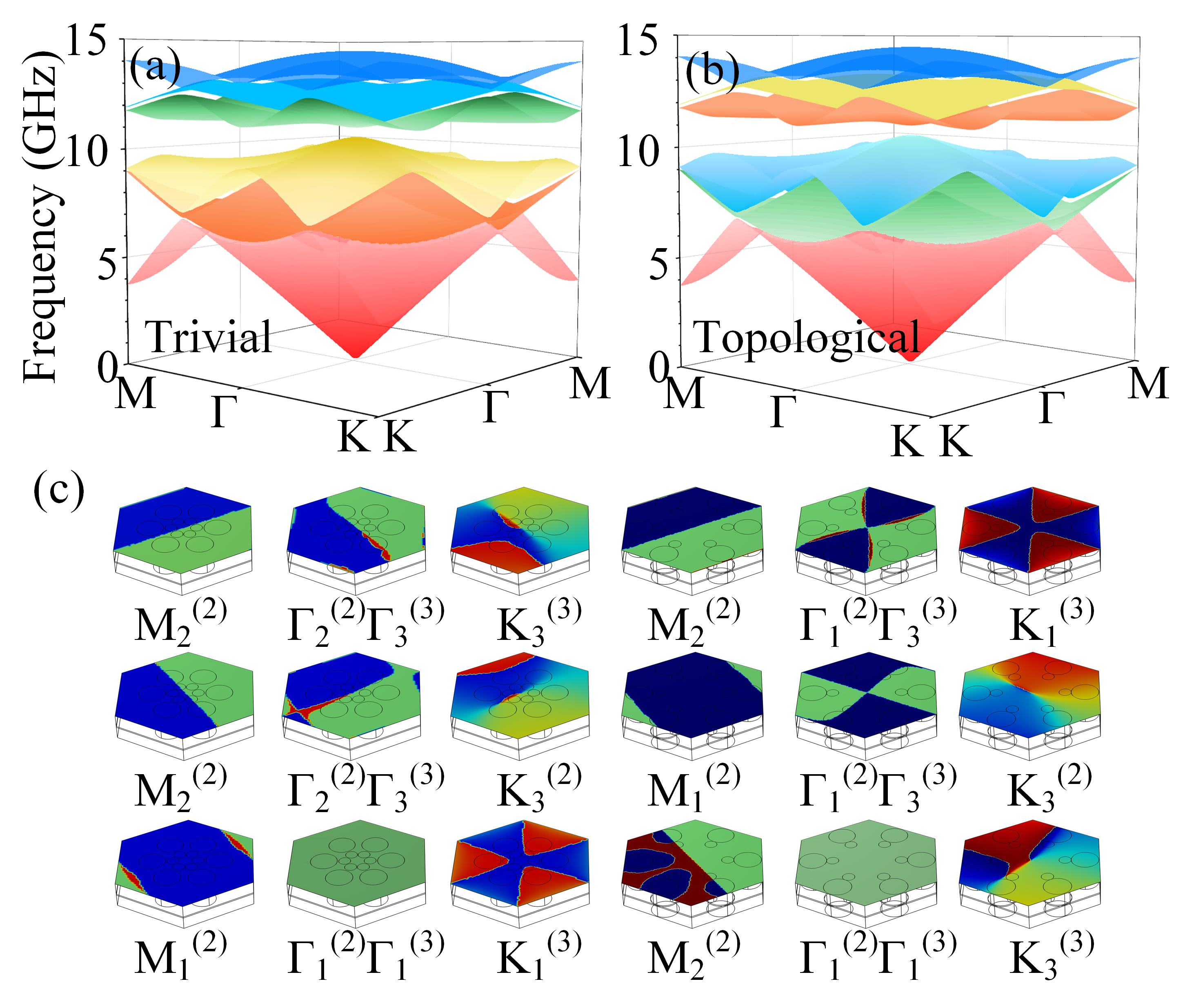}\\
   \caption{\label{Fig.4}Under the conditions $a=18mm,r_1=1.8mm,r_2=0.8mm,h=0.1mm,b=0.11a~mm$: (a) Trivial states: $\beta=2.6/a$ corresponding to $l_1=4.44mm, l_2=1.7mm$. (b) Topological states: $\beta=2.74/a$ corresponding to $l_1=7.1mm, l_2=4.5mm$. (c) Phase distribution from high-symmetry point M to  $\Gamma$ and then to K: The left and right sections correspond to Figs.~\ref{Fig.4}(a) and ~\ref{Fig.4}(b), respectively. }
\end{figure}

 Choosing the TNPC with $\varphi =$60$^\circ$  for investigation, its band structure and phase distribution are illustrated in Fig.~\ref{Fig.4}.
The band structures of the TNPC with two different $\beta$ are shown in Figs. ~\ref{Fig.4}(a)-\ref{Fig.4}(b). Since the upper and lower layers are identical PCs, $\beta_1=\beta_2=\beta$ can be set. Regardless of how the nested factor changes, $C_\mathrm6$ symmetry is always satisfied. Therefore, the topological properties can be determined by calculating the topological index. The corresponding eigenvalues of the rotation symmetry at the high-symmetry points below the bandgap are given by $\Pi_p^{(n)}=e^{2 \pi i(p-1) / n}$~\cite{RN34,RN35} ， $p$ represents the eigenvalue index, $p =1,2,3,..., n$.The Brillouin zone has three high-symmetry points, M, $\Gamma$, and K, which are isomorphic to  $C_\mathrm2$   ,  $C_\mathrm6$   , and $C_\mathrm3$    symmetries, respectively. Therefore, the irreducible representations at the high-symmetry points can be determined using the corresponding point group character table \cite{RN36}. As shown in Fig. ~\ref{Fig.4}(c), the eigenvalues of rotation symmetry for the nested factor are described sequentially from M to $\Gamma$ and then to K, 
\begin{equation}
\begin{aligned}
		\label{eq.14}
		\beta=2.6 / a,\{&\mathrm{M}_1^{(2)},\mathrm{M}_2^{(2)}, \mathrm{M}_2^{(2)} ; \mathrm{\Gamma}_1^{(2)} \mathrm{\Gamma}_1^{(3)},  \mathrm{\Gamma}_2^{(2)} \mathrm{\Gamma}_3^{(3)},\\
       &\mathrm{\Gamma}_2^{(2)}\mathrm{\Gamma}_3^{(3)} ;
       \mathrm{K}_3^{(3)},\mathrm{K}_3^{(3)},\mathrm{K}_1^{(3)}\}
       \end{aligned}
	\end{equation}
 \begin{equation}
 \begin{aligned}
		\label{eq.15}
		\beta=2.74 / a,\{&\mathrm{M}_1^{(2)},\mathrm{M}_2^{(2)}, \mathrm{M}_2^{(2)} ; \mathrm{\Gamma}_1^{(2)} \mathrm{\Gamma}_1^{(3)},  \mathrm{\Gamma}_1^{(2)} \mathrm{\Gamma}_3^{(3)},\\
  &\mathrm{\Gamma}_1^{(2)}\mathrm{\Gamma}_3^{(3)} ; \mathrm{K}_3^{(3)},\mathrm{K}_3^{(2)},\mathrm{K}_1^{(3)}\}
       \end{aligned}
	\end{equation}
 
 Previous researches have provided a significant theoretical basis and a classification framework for the study of topological physics by systematically identifying and classifying the topological states present in different crystal structures, through considering the space group symmetries of crystals~\cite{RN37,RN38}. Defining whether a state is topological can be achieved by comparing the differences in wave functions at different high-symmetry points. Therefore, a new symmetry-indicator topological invariant can be defined by comparing the rotation eigenvalues at high-symmetry point $\Pi_p^{(n)}$ with those at a reference point  $\Gamma_p^{(n)}$ \cite{RN36,RN37,RN38,RN39}:$\left[\Pi_p^{(n)}\right]=\# \Pi_p^{(n)}-\# \Gamma_p^{(n)}$ , where $\# \Pi_p^{(n)}$ represents the number of bands with $C_n$ rotation symmetry eigenvalues at the high-symmetry point $\Pi=\Gamma$,~K,~M  below the bandgap. Consequently, the topological indicator $\chi^{(6)}=\left(\left[\mathrm{M}_1^{(2)}\right],\left[\mathrm{K}_1^{(3)}\right]\right)$can be ascertained to determine the presence of a topological state in the structure with $C_\mathrm6$ symmetry \cite{RN37,RN38,RN39,RN40,RN41}.
 
When \(\beta=2.6/a\), \(\chi^{(6)}=(0,0)\). When \(\beta=2.74/a\), \(\chi^{(6)}=(-2,0)\), the non-zero topological index indicates that the system is in a topological state. Moreover, upon reviewing the point group table, the corresponding irreducible representation is given by,
 \begin{equation}
      \label{eq.16}
\beta=2.6/ a,\left\{A, B, B ; A, E_2, E_2 ;A, E, E\right\}
 \end{equation}
 \vspace{-\baselineskip} 
 \begin{equation}
  \label{eq.17}
 \beta=2.74/ a,\left\{B, A, B ; A, E_1, E_1 ; E, E, A\right\}
  \end{equation}
  
The rotation eigenvalues of the TNPC can be determined by combining the phase distribution with the point group character table. Upon conversion to the H-M symbol, it is revealed that this structure belongs to the $P_6$ space group, with the most significant Wyckoff positions being 1a, 2b, and 3c. By consulting the Bilbao crystallographic server, it is determined that under the $P_6$ space group, the fundamental band representation (EBR) is available, as detailed in Sec. D in Supplemental Material. This enables the identification of the Wannier centers corresponding to each band, with the analysis of degenerate bands requiring a holistic approach. Consequently, the relationship between all bands below the gap and EBR can be established: $\text { i) }$ When $\beta=2.6/a$, The EBR for the first band is denoted as $\left(A\uparrow G\right)_{1a}$, while that for the second and third bands is represented by $\left({ }^1 \mathrm{E}_2{ }^2 \mathrm{E}_2 \uparrow \mathrm{G}\right)_{1 a}$. Therefore, all the Wannier centers corresponding to the bands below the gap are located at 1a, precisely at the center of the unit cell. In this case, the system cannot generate topological edge states or HOTSs, indicating that it is in a trivial states. $\text { ii) }$ When $\beta=2.74/a$, since the three bands degenerate, EBR for the three bands is denoted as $\left(A\uparrow G\right)_{3c}$. Therefore, all Wannier centers associated with bands below the gap are found at 3c, precisely at he central position on the edge of the unit cell. In this configuration, topological edge states and higher-order topological states can emerge, indicating that TNPC is in topological states. 

\begin{figure} 
		\centerline{\includegraphics[scale=.5]{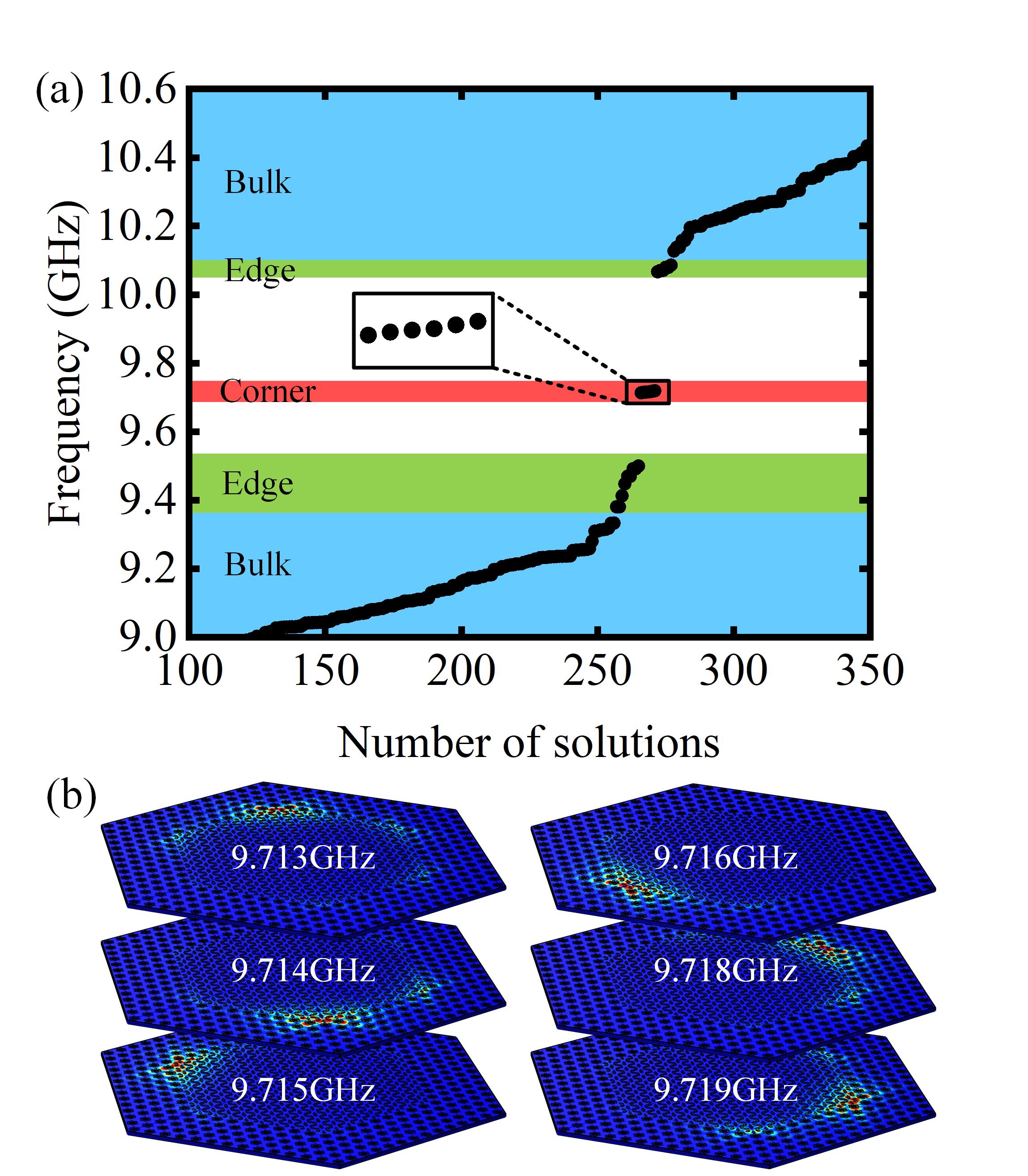}}
		\caption{\label{Fig.5}(a) Eigenmodes of a box-shaped structure. (b) Six-fold degenerate HOTSs with $C_\mathrm{2z}$ symmetry.}
	\end{figure}

Furthermore, based on the box-shaped structure, the distribution of its eigenfrequencies and electric field is illustrated in Fig. ~\ref{Fig.5}. 
Since the TNPC still maintains $C_\mathrm6$  symmetry, the existence of HOTSs can be determined by calculating whether the quadrupole moment is nonzero~\cite{RN40}:
 \begin{equation}
  \label{eq.18}
Q_{c}^{(6)} = \frac{1}{4}[\mathrm{M_1^{(2)}}]+\frac{1}{6}[K_1^{(3)}]\mathrm{mod1}
\end{equation}

According to Eq.~\eqref{eq.18}, $Q_{c}^{(6)}=1/4$ . Therefore, the system exhibits HOTSs. From Fig.  ~\ref{Fig.5}(a), it is evident that the TNPC possesses six-fold degenerate HOTSs with  $C_\mathrm{2z}$ symmetry. And the field distribution is illustrated in Fig.  ~\ref{Fig.5}(b). Research indicates that by tuning the twisting and nesting, there is a potential to further adjust the positions, intensities, and mode numbers of these HOTSs. This insight can serve as a theoretical reference for the design of relevant topological photonic devices. 

 The SGRs and the aperiodic-quasiperiodic-periodic properties in twisted  nested Moir\texorpdfstring{$\mathbf{\acute{e}}$}{e} patterns  were derived. Additionally, the SSH\(\varphi\) model Hamiltonian was established, revealing the intrinsic connection between twisted  nested Moir\texorpdfstring{$\mathbf{\acute{e}}$}{e} patterns and topological transitions. Furthermore, HOTSs with \(C_{2z}\) symmetry were achieved. The integration of the TNPC contributes to a deeper understanding of complex optical properties, providing a new theoretical framework for the study of topological waveguide transmission and localization.
 
\addtolength{\parskip}{0.5em}
This work was supported by the National Natural Science Foundation of China (Grants No. 61405058 and No. 62075059), the Open Project of State Key Laboratory of Advanced Optical Communication Systems and Networks of China (Grant No. 2024GZKF20), the Natural Science Foundation of Hunan Province (Grants No. 2017JJ2048 and No. 2020JJ4161), and the Scientific Research Foundation of Hunan Provincal Education Department (Grant No. 21A0013).
\nocite{*}

\begin{thebibliography}{41}%
\makeatletter
\providecommand \@ifxundefined [1]{%
 \@ifx{#1\undefined}
}%
\providecommand \@ifnum [1]{%
 \ifnum #1\expandafter \@firstoftwo
 \else \expandafter \@secondoftwo
 \fi
}%
\providecommand \@ifx [1]{%
 \ifx #1\expandafter \@firstoftwo
 \else \expandafter \@secondoftwo
 \fi
}%
\providecommand \natexlab [1]{#1}%
\providecommand \enquote  [1]{``#1''}%
\providecommand \bibnamefont  [1]{#1}%
\providecommand \bibfnamefont [1]{#1}%
\providecommand \citenamefont [1]{#1}%
\providecommand \href@noop [0]{\@secondoftwo}%
\providecommand \href [0]{\begingroup \@sanitize@url \@href}%
\providecommand \@href[1]{\@@startlink{#1}\@@href}%
\providecommand \@@href[1]{\endgroup#1\@@endlink}%
\providecommand \@sanitize@url [0]{\catcode `\\12\catcode `\$12\catcode
  `\&12\catcode `\#12\catcode `\^12\catcode `\_12\catcode `\%12\relax}%
\providecommand \@@startlink[1]{}%
\providecommand \@@endlink[0]{}%
\providecommand \url  [0]{\begingroup\@sanitize@url \@url }%
\providecommand \@url [1]{\endgroup\@href {#1}{\urlprefix }}%
\providecommand \urlprefix  [0]{URL }%
\providecommand \Eprint [0]{\href }%
\providecommand \doibase [0]{https://doi.org/}%
\providecommand \selectlanguage [0]{\@gobble}%
\providecommand \bibinfo  [0]{\@secondoftwo}%
\providecommand \bibfield  [0]{\@secondoftwo}%
\providecommand \translation [1]{[#1]}%
\providecommand \BibitemOpen [0]{}%
\providecommand \bibitemStop [0]{}%
\providecommand \bibitemNoStop [0]{.\EOS\space}%
\providecommand \EOS [0]{\spacefactor3000\relax}%
\providecommand \BibitemShut  [1]{\csname bibitem#1\endcsname}%
\let\auto@bib@innerbib\@empty
\bibitem [{\citenamefont {Chen}\ \emph {et~al.}(2021)\citenamefont {Chen},
  \citenamefont {Lin}, \citenamefont {Chen}, \citenamefont {Low}, \citenamefont
  {Chen},\ and\ \citenamefont {Dai}}]{RN1}%
  \BibitemOpen
  \bibfield  {author} {\bibinfo {author} {\bibfnamefont {J.}~\bibnamefont
  {Chen}}, \bibinfo {author} {\bibfnamefont {X.}~\bibnamefont {Lin}}, \bibinfo
  {author} {\bibfnamefont {M.}~\bibnamefont {Chen}}, \bibinfo {author}
  {\bibfnamefont {T.}~\bibnamefont {Low}}, \bibinfo {author} {\bibfnamefont
  {H.}~\bibnamefont {Chen}},\ and\ \bibinfo {author} {\bibfnamefont
  {S.}~\bibnamefont {Dai}},\ }\bibfield  {title} {\bibinfo {title} {{A
  perspective of twisted photonic structures}},\ }\href
  {https://doi.org/10.1063/5.0070163} {\bibfield  {journal} {\bibinfo
  {journal} {Appl. Phys. Lett.}\ }\textbf {\bibinfo {volume} {119}},\ \bibinfo
  {pages} {240501} (\bibinfo {year} {2021})}\BibitemShut {NoStop}%
\bibitem [{\citenamefont {Wang}\ \emph {et~al.}(2018)\citenamefont {Wang},
  \citenamefont {Gupta}, \citenamefont {Zhu}, \citenamefont {Lu}, \citenamefont
  {Liu},\ and\ \citenamefont {Chen}}]{RN2}%
  \BibitemOpen
  \bibfield  {author} {\bibinfo {author} {\bibfnamefont {H.-F.}\ \bibnamefont
  {Wang}}, \bibinfo {author} {\bibfnamefont {S.~K.}\ \bibnamefont {Gupta}},
  \bibinfo {author} {\bibfnamefont {X.-Y.}\ \bibnamefont {Zhu}}, \bibinfo
  {author} {\bibfnamefont {M.-H.}\ \bibnamefont {Lu}}, \bibinfo {author}
  {\bibfnamefont {X.-P.}\ \bibnamefont {Liu}},\ and\ \bibinfo {author}
  {\bibfnamefont {Y.-F.}\ \bibnamefont {Chen}},\ }\bibfield  {title} {\bibinfo
  {title} {Bound states in the continuum in a bilayer photonic crystal with
  te-tm cross coupling},\ }\href {https://doi.org/10.1103/PhysRevB.98.214101}
  {\bibfield  {journal} {\bibinfo  {journal} {Phys. Rev. B}\ }\textbf {\bibinfo
  {volume} {98}},\ \bibinfo {pages} {214101} (\bibinfo {year}
  {2018})}\BibitemShut {NoStop}%
\bibitem [{\citenamefont {Huang}\ \emph {et~al.}(2022)\citenamefont {Huang},
  \citenamefont {Zhang},\ and\ \citenamefont {Zhang}}]{RN3}%
  \BibitemOpen
  \bibfield  {author} {\bibinfo {author} {\bibfnamefont {L.}~\bibnamefont
  {Huang}}, \bibinfo {author} {\bibfnamefont {W.}~\bibnamefont {Zhang}},\ and\
  \bibinfo {author} {\bibfnamefont {X.}~\bibnamefont {Zhang}},\ }\bibfield
  {title} {\bibinfo {title} {Moir\'e quasibound states in the continuum},\
  }\href {https://doi.org/10.1103/PhysRevLett.128.253901} {\bibfield  {journal}
  {\bibinfo  {journal} {Phys. Rev. Lett.}\ }\textbf {\bibinfo {volume} {128}},\
  \bibinfo {pages} {253901} (\bibinfo {year} {2022})}\BibitemShut {NoStop}%
\bibitem [{\citenamefont {Chen}\ \emph {et~al.}(2019)\citenamefont {Chen},
  \citenamefont {He},\ and\ \citenamefont {Dong}}]{RN4}%
  \BibitemOpen
  \bibfield  {author} {\bibinfo {author} {\bibfnamefont {X.-D.}\ \bibnamefont
  {Chen}}, \bibinfo {author} {\bibfnamefont {X.-T.}\ \bibnamefont {He}},\ and\
  \bibinfo {author} {\bibfnamefont {J.-W.}\ \bibnamefont {Dong}},\ }\bibfield
  {title} {\bibinfo {title} {All-dielectric layered photonic topological
  insulators},\ }\href {https://doi.org/10.1002/lpor.201900091} {\bibfield
  {journal} {\bibinfo  {journal} {Laser Photonics Rev.}\ }\textbf {\bibinfo
  {volume} {13}},\ \bibinfo {pages} {1900091} (\bibinfo {year}
  {2019})}\BibitemShut {NoStop}%
\bibitem [{\citenamefont {Qi}\ \emph {et~al.}(2019)\citenamefont {Qi},
  \citenamefont {Niu}, \citenamefont {Zhang}, \citenamefont {Wu}, \citenamefont
  {Chu}, \citenamefont {Ma},\ and\ \citenamefont {Tang}}]{RN5}%
  \BibitemOpen
  \bibfield  {author} {\bibinfo {author} {\bibfnamefont {Y.}~\bibnamefont
  {Qi}}, \bibinfo {author} {\bibfnamefont {W.}~\bibnamefont {Niu}}, \bibinfo
  {author} {\bibfnamefont {S.}~\bibnamefont {Zhang}}, \bibinfo {author}
  {\bibfnamefont {S.}~\bibnamefont {Wu}}, \bibinfo {author} {\bibfnamefont
  {L.}~\bibnamefont {Chu}}, \bibinfo {author} {\bibfnamefont {W.}~\bibnamefont
  {Ma}},\ and\ \bibinfo {author} {\bibfnamefont {B.}~\bibnamefont {Tang}},\
  }\bibfield  {title} {\bibinfo {title} {Encoding and decoding of invisible
  complex information in a dual-response bilayer photonic crystal with tunable
  wettability},\ }\href {https://doi.org/10.1002/adfm.201906799} {\bibfield
  {journal} {\bibinfo  {journal} {Adv. Funct. Mater.}\ }\textbf {\bibinfo
  {volume} {29}},\ \bibinfo {pages} {1906799} (\bibinfo {year}
  {2019})}\BibitemShut {NoStop}%
\bibitem [{\citenamefont {Wang}\ \emph
  {et~al.}(2020{\natexlab{a}})\citenamefont {Wang}, \citenamefont {Zheng},
  \citenamefont {Chen}, \citenamefont {Huang}, \citenamefont {Kartashov},
  \citenamefont {Torner}, \citenamefont {Konotop},\ and\ \citenamefont
  {Ye}}]{RN6}%
  \BibitemOpen
  \bibfield  {author} {\bibinfo {author} {\bibfnamefont {P.}~\bibnamefont
  {Wang}}, \bibinfo {author} {\bibfnamefont {Y.}~\bibnamefont {Zheng}},
  \bibinfo {author} {\bibfnamefont {X.}~\bibnamefont {Chen}}, \bibinfo {author}
  {\bibfnamefont {C.}~\bibnamefont {Huang}}, \bibinfo {author} {\bibfnamefont
  {Y.~V.}\ \bibnamefont {Kartashov}}, \bibinfo {author} {\bibfnamefont
  {L.}~\bibnamefont {Torner}}, \bibinfo {author} {\bibfnamefont {V.~V.}\
  \bibnamefont {Konotop}},\ and\ \bibinfo {author} {\bibfnamefont
  {F.}~\bibnamefont {Ye}},\ }\bibfield  {title} {\bibinfo {title} {Localization
  and delocalization of light in photonic moir{\'e} lattices},\ }\href
  {https://doi.org/10.1038/s41586-019-1851-6} {\bibfield  {journal} {\bibinfo
  {journal} {Nature}\ }\textbf {\bibinfo {volume} {577}},\ \bibinfo {pages}
  {42} (\bibinfo {year} {2020}{\natexlab{a}})}\BibitemShut {NoStop}%
\bibitem [{\citenamefont {Mao}\ \emph {et~al.}(2021)\citenamefont {Mao},
  \citenamefont {Shao}, \citenamefont {Luan}, \citenamefont {Wang},\ and\
  \citenamefont {Ma}}]{RN7}%
  \BibitemOpen
  \bibfield  {author} {\bibinfo {author} {\bibfnamefont {X.-R.}\ \bibnamefont
  {Mao}}, \bibinfo {author} {\bibfnamefont {Z.-K.}\ \bibnamefont {Shao}},
  \bibinfo {author} {\bibfnamefont {H.-Y.}\ \bibnamefont {Luan}}, \bibinfo
  {author} {\bibfnamefont {S.-L.}\ \bibnamefont {Wang}},\ and\ \bibinfo
  {author} {\bibfnamefont {R.-M.}\ \bibnamefont {Ma}},\ }\bibfield  {title}
  {\bibinfo {title} {Magic-angle lasers in nanostructured moir{\'e}
  superlattice},\ }\href {https://doi.org/10.1038/s41565-021-00956-7}
  {\bibfield  {journal} {\bibinfo  {journal} {Nat. Nanotechnol.}\ }\textbf
  {\bibinfo {volume} {16}},\ \bibinfo {pages} {1099} (\bibinfo {year}
  {2021})}\BibitemShut {NoStop}%
\bibitem [{\citenamefont {Luan}\ \emph {et~al.}(2023)\citenamefont {Luan},
  \citenamefont {Ouyang}, \citenamefont {Zhao}, \citenamefont {Mao},\ and\
  \citenamefont {Ma}}]{RN8}%
  \BibitemOpen
  \bibfield  {author} {\bibinfo {author} {\bibfnamefont {H.-Y.}\ \bibnamefont
  {Luan}}, \bibinfo {author} {\bibfnamefont {Y.-H.}\ \bibnamefont {Ouyang}},
  \bibinfo {author} {\bibfnamefont {Z.-W.}\ \bibnamefont {Zhao}}, \bibinfo
  {author} {\bibfnamefont {W.-Z.}\ \bibnamefont {Mao}},\ and\ \bibinfo {author}
  {\bibfnamefont {R.-M.}\ \bibnamefont {Ma}},\ }\bibfield  {title} {\bibinfo
  {title} {Reconfigurable moir{\'e} nanolaser arrays with phase
  synchronization},\ }\href {https://doi.org/10.1038/s41586-023-06789-9}
  {\bibfield  {journal} {\bibinfo  {journal} {Nature}\ }\textbf {\bibinfo
  {volume} {624}},\ \bibinfo {pages} {282} (\bibinfo {year}
  {2023})}\BibitemShut {NoStop}%
\bibitem [{\citenamefont {Wang}\ \emph
  {et~al.}(2020{\natexlab{b}})\citenamefont {Wang}, \citenamefont {Gao},
  \citenamefont {Chen}, \citenamefont {Shi}, \citenamefont {Li}, \citenamefont
  {Dong}, \citenamefont {Xiang},\ and\ \citenamefont {Zhang}}]{RN9}%
  \BibitemOpen
  \bibfield  {author} {\bibinfo {author} {\bibfnamefont {W.}~\bibnamefont
  {Wang}}, \bibinfo {author} {\bibfnamefont {W.}~\bibnamefont {Gao}}, \bibinfo
  {author} {\bibfnamefont {X.}~\bibnamefont {Chen}}, \bibinfo {author}
  {\bibfnamefont {F.}~\bibnamefont {Shi}}, \bibinfo {author} {\bibfnamefont
  {G.}~\bibnamefont {Li}}, \bibinfo {author} {\bibfnamefont {J.}~\bibnamefont
  {Dong}}, \bibinfo {author} {\bibfnamefont {Y.}~\bibnamefont {Xiang}},\ and\
  \bibinfo {author} {\bibfnamefont {S.}~\bibnamefont {Zhang}},\ }\bibfield
  {title} {\bibinfo {title} {Moir\'e fringe induced gauge field in photonics},\
  }\href {https://doi.org/10.1103/PhysRevLett.125.203901} {\bibfield  {journal}
  {\bibinfo  {journal} {Phys. Rev. Lett.}\ }\textbf {\bibinfo {volume} {125}},\
  \bibinfo {pages} {203901} (\bibinfo {year} {2020}{\natexlab{b}})}\BibitemShut
  {NoStop}%
\bibitem [{\citenamefont {Tang}\ \emph {et~al.}(2021)\citenamefont {Tang},
  \citenamefont {Du}, \citenamefont {Carr}, \citenamefont {DeVault},
  \citenamefont {Mello},\ and\ \citenamefont {Mazur}}]{RN10}%
  \BibitemOpen
  \bibfield  {author} {\bibinfo {author} {\bibfnamefont {H.}~\bibnamefont
  {Tang}}, \bibinfo {author} {\bibfnamefont {F.}~\bibnamefont {Du}}, \bibinfo
  {author} {\bibfnamefont {S.}~\bibnamefont {Carr}}, \bibinfo {author}
  {\bibfnamefont {C.}~\bibnamefont {DeVault}}, \bibinfo {author} {\bibfnamefont
  {O.}~\bibnamefont {Mello}},\ and\ \bibinfo {author} {\bibfnamefont
  {E.}~\bibnamefont {Mazur}},\ }\bibfield  {title} {\bibinfo {title} {Modeling
  the optical properties of twisted bilayer photonic crystals},\ }\href
  {https://doi.org/10.1038/s41377-021-00601-x} {\bibfield  {journal} {\bibinfo
  {journal} {Light Sci. Appl.}\ }\textbf {\bibinfo {volume} {10}},\ \bibinfo
  {pages} {157} (\bibinfo {year} {2021})}\BibitemShut {NoStop}%
\bibitem [{\citenamefont {Nguyen}\ \emph {et~al.}(2022)\citenamefont {Nguyen},
  \citenamefont {Letartre}, \citenamefont {Drouard}, \citenamefont
  {Viktorovitch}, \citenamefont {Nguyen},\ and\ \citenamefont {Nguyen}}]{RN11}%
  \BibitemOpen
  \bibfield  {author} {\bibinfo {author} {\bibfnamefont {D.~X.}\ \bibnamefont
  {Nguyen}}, \bibinfo {author} {\bibfnamefont {X.}~\bibnamefont {Letartre}},
  \bibinfo {author} {\bibfnamefont {E.}~\bibnamefont {Drouard}}, \bibinfo
  {author} {\bibfnamefont {P.}~\bibnamefont {Viktorovitch}}, \bibinfo {author}
  {\bibfnamefont {H.~C.}\ \bibnamefont {Nguyen}},\ and\ \bibinfo {author}
  {\bibfnamefont {H.~S.}\ \bibnamefont {Nguyen}},\ }\bibfield  {title}
  {\bibinfo {title} {Magic configurations in moir\'e superlattice of bilayer
  photonic crystals: Almost-perfect flatbands and unconventional
  localization},\ }\href {https://doi.org/10.1103/PhysRevResearch.4.L032031}
  {\bibfield  {journal} {\bibinfo  {journal} {Phys. Rev. Res.}\ }\textbf
  {\bibinfo {volume} {4}},\ \bibinfo {pages} {L032031} (\bibinfo {year}
  {2022})}\BibitemShut {NoStop}%
\bibitem [{\citenamefont {Lou}\ \emph {et~al.}(2021)\citenamefont {Lou},
  \citenamefont {Zhao}, \citenamefont {Minkov}, \citenamefont {Guo},
  \citenamefont {Orenstein},\ and\ \citenamefont {Fan}}]{RN12}%
  \BibitemOpen
  \bibfield  {author} {\bibinfo {author} {\bibfnamefont {B.}~\bibnamefont
  {Lou}}, \bibinfo {author} {\bibfnamefont {N.}~\bibnamefont {Zhao}}, \bibinfo
  {author} {\bibfnamefont {M.}~\bibnamefont {Minkov}}, \bibinfo {author}
  {\bibfnamefont {C.}~\bibnamefont {Guo}}, \bibinfo {author} {\bibfnamefont
  {M.}~\bibnamefont {Orenstein}},\ and\ \bibinfo {author} {\bibfnamefont
  {S.}~\bibnamefont {Fan}},\ }\bibfield  {title} {\bibinfo {title} {Theory for
  twisted bilayer photonic crystal slabs},\ }\href
  {https://doi.org/10.1103/PhysRevLett.126.136101} {\bibfield  {journal}
  {\bibinfo  {journal} {Phys. Rev. Lett.}\ }\textbf {\bibinfo {volume} {126}},\
  \bibinfo {pages} {136101} (\bibinfo {year} {2021})}\BibitemShut {NoStop}%
\bibitem [{\citenamefont {Zhou}\ \emph {et~al.}(2020)\citenamefont {Zhou},
  \citenamefont {Lin}, \citenamefont {Lu}, \citenamefont {Lai}, \citenamefont
  {Hou},\ and\ \citenamefont {Jiang}}]{RN13}%
  \BibitemOpen
  \bibfield  {author} {\bibinfo {author} {\bibfnamefont {X.}~\bibnamefont
  {Zhou}}, \bibinfo {author} {\bibfnamefont {Z.-K.}\ \bibnamefont {Lin}},
  \bibinfo {author} {\bibfnamefont {W.}~\bibnamefont {Lu}}, \bibinfo {author}
  {\bibfnamefont {Y.}~\bibnamefont {Lai}}, \bibinfo {author} {\bibfnamefont
  {B.}~\bibnamefont {Hou}},\ and\ \bibinfo {author} {\bibfnamefont {J.-H.}\
  \bibnamefont {Jiang}},\ }\bibfield  {title} {\bibinfo {title} {Twisted
  quadrupole topological photonic crystals},\ }\href
  {https://doi.org/https://doi.org/10.1002/lpor.202000010} {\bibfield
  {journal} {\bibinfo  {journal} {Laser Photonics Rev.}\ }\textbf {\bibinfo
  {volume} {14}},\ \bibinfo {pages} {2000010} (\bibinfo {year}
  {2020})}\BibitemShut {NoStop}%
\bibitem [{\citenamefont {Oudich}\ \emph {et~al.}(2021)\citenamefont {Oudich},
  \citenamefont {Su}, \citenamefont {Deng}, \citenamefont {Benalcazar},
  \citenamefont {Huang}, \citenamefont {Gerard}, \citenamefont {Lu},
  \citenamefont {Zhan},\ and\ \citenamefont {Jing}}]{RN14}%
  \BibitemOpen
  \bibfield  {author} {\bibinfo {author} {\bibfnamefont {M.}~\bibnamefont
  {Oudich}}, \bibinfo {author} {\bibfnamefont {G.}~\bibnamefont {Su}}, \bibinfo
  {author} {\bibfnamefont {Y.}~\bibnamefont {Deng}}, \bibinfo {author}
  {\bibfnamefont {W.}~\bibnamefont {Benalcazar}}, \bibinfo {author}
  {\bibfnamefont {R.}~\bibnamefont {Huang}}, \bibinfo {author} {\bibfnamefont
  {N.~J. R.~K.}\ \bibnamefont {Gerard}}, \bibinfo {author} {\bibfnamefont
  {M.}~\bibnamefont {Lu}}, \bibinfo {author} {\bibfnamefont {P.}~\bibnamefont
  {Zhan}},\ and\ \bibinfo {author} {\bibfnamefont {Y.}~\bibnamefont {Jing}},\
  }\bibfield  {title} {\bibinfo {title} {Photonic analog of bilayer graphene},\
  }\href {https://doi.org/10.1103/PhysRevB.103.214311} {\bibfield  {journal}
  {\bibinfo  {journal} {Phys. Rev. B}\ }\textbf {\bibinfo {volume} {103}},\
  \bibinfo {pages} {214311} (\bibinfo {year} {2021})}\BibitemShut {NoStop}%
\bibitem [{\citenamefont {Zhang}\ \emph
  {et~al.}(2022{\natexlab{a}})\citenamefont {Zhang}, \citenamefont {Tang},
  \citenamefont {Dai}, \citenamefont {Zhang},\ and\ \citenamefont
  {Xiang}}]{RN15}%
  \BibitemOpen
  \bibfield  {author} {\bibinfo {author} {\bibfnamefont {Y.}~\bibnamefont
  {Zhang}}, \bibinfo {author} {\bibfnamefont {J.}~\bibnamefont {Tang}},
  \bibinfo {author} {\bibfnamefont {X.}~\bibnamefont {Dai}}, \bibinfo {author}
  {\bibfnamefont {S.}~\bibnamefont {Zhang}},\ and\ \bibinfo {author}
  {\bibfnamefont {Y.}~\bibnamefont {Xiang}},\ }\bibfield  {title} {\bibinfo
  {title} {Higher-order nodal ring photonic semimetal},\ }\href
  {https://doi.org/10.1364/OL.472397} {\bibfield  {journal} {\bibinfo
  {journal} {Opt. Lett.}\ }\textbf {\bibinfo {volume} {47}},\ \bibinfo {pages}
  {5885} (\bibinfo {year} {2022}{\natexlab{a}})}\BibitemShut {NoStop}%
\bibitem [{\citenamefont {He}\ \emph {et~al.}(2021)\citenamefont {He},
  \citenamefont {Zhou}, \citenamefont {Ye}, \citenamefont {Cho}, \citenamefont
  {Jeong}, \citenamefont {Meng},\ and\ \citenamefont {Wang}}]{RN16}%
  \BibitemOpen
  \bibfield  {author} {\bibinfo {author} {\bibfnamefont {F.}~\bibnamefont
  {He}}, \bibinfo {author} {\bibfnamefont {Y.}~\bibnamefont {Zhou}}, \bibinfo
  {author} {\bibfnamefont {Z.}~\bibnamefont {Ye}}, \bibinfo {author}
  {\bibfnamefont {S.-H.}\ \bibnamefont {Cho}}, \bibinfo {author} {\bibfnamefont
  {J.}~\bibnamefont {Jeong}}, \bibinfo {author} {\bibfnamefont
  {X.}~\bibnamefont {Meng}},\ and\ \bibinfo {author} {\bibfnamefont
  {Y.}~\bibnamefont {Wang}},\ }\bibfield  {title} {\bibinfo {title} {Moir{\'e}
  patterns in 2d materials: A review},\ }\href
  {https://doi.org/10.1021/acsnano.0c10435} {\bibfield  {journal} {\bibinfo
  {journal} {ACS nano}\ }\textbf {\bibinfo {volume} {15}},\ \bibinfo {pages}
  {5944} (\bibinfo {year} {2021})}\BibitemShut {NoStop}%
\bibitem [{\citenamefont {Alden}\ \emph {et~al.}(2013)\citenamefont {Alden},
  \citenamefont {Tsen}, \citenamefont {Huang}, \citenamefont {Hovden},
  \citenamefont {Brown}, \citenamefont {Park}, \citenamefont {Muller},\ and\
  \citenamefont {McEuen}}]{RN17}%
  \BibitemOpen
  \bibfield  {author} {\bibinfo {author} {\bibfnamefont {J.~S.}\ \bibnamefont
  {Alden}}, \bibinfo {author} {\bibfnamefont {A.~W.}\ \bibnamefont {Tsen}},
  \bibinfo {author} {\bibfnamefont {P.~Y.}\ \bibnamefont {Huang}}, \bibinfo
  {author} {\bibfnamefont {R.}~\bibnamefont {Hovden}}, \bibinfo {author}
  {\bibfnamefont {L.}~\bibnamefont {Brown}}, \bibinfo {author} {\bibfnamefont
  {J.}~\bibnamefont {Park}}, \bibinfo {author} {\bibfnamefont {D.~A.}\
  \bibnamefont {Muller}},\ and\ \bibinfo {author} {\bibfnamefont {P.~L.}\
  \bibnamefont {McEuen}},\ }\bibfield  {title} {\bibinfo {title} {Strain
  solitons and topological defects in bilayer graphene},\ }\href
  {https://doi.org/10.1073/pnas.130939411} {\bibfield  {journal} {\bibinfo
  {journal} {Proc. Natl. Acad. Sci.}\ }\textbf {\bibinfo {volume} {110}},\
  \bibinfo {pages} {11256} (\bibinfo {year} {2013})}\BibitemShut {NoStop}%
\bibitem [{\citenamefont {Lin}\ \emph {et~al.}(2013)\citenamefont {Lin},
  \citenamefont {Fang}, \citenamefont {Zhou}, \citenamefont {Lupini},
  \citenamefont {Idrobo}, \citenamefont {Kong}, \citenamefont {Pennycook},\
  and\ \citenamefont {Pantelides}}]{RN18}%
  \BibitemOpen
  \bibfield  {author} {\bibinfo {author} {\bibfnamefont {J.}~\bibnamefont
  {Lin}}, \bibinfo {author} {\bibfnamefont {W.}~\bibnamefont {Fang}}, \bibinfo
  {author} {\bibfnamefont {W.}~\bibnamefont {Zhou}}, \bibinfo {author}
  {\bibfnamefont {A.~R.}\ \bibnamefont {Lupini}}, \bibinfo {author}
  {\bibfnamefont {J.~C.}\ \bibnamefont {Idrobo}}, \bibinfo {author}
  {\bibfnamefont {J.}~\bibnamefont {Kong}}, \bibinfo {author} {\bibfnamefont
  {S.~J.}\ \bibnamefont {Pennycook}},\ and\ \bibinfo {author} {\bibfnamefont
  {S.~T.}\ \bibnamefont {Pantelides}},\ }\bibfield  {title} {\bibinfo {title}
  {Ac/ab stacking boundaries in bilayer graphene},\ }\href
  {https://doi.org/10.1021/nl4013979} {\bibfield  {journal} {\bibinfo
  {journal} {Nano Lett.}\ }\textbf {\bibinfo {volume} {13}},\ \bibinfo {pages}
  {3262} (\bibinfo {year} {2013})}\BibitemShut {NoStop}%
\bibitem [{\citenamefont {Ni}\ \emph {et~al.}(2015)\citenamefont {Ni},
  \citenamefont {Wang}, \citenamefont {Wu}, \citenamefont {Fei}, \citenamefont
  {Goldflam}, \citenamefont {Keilmann}, \citenamefont {{\"O}zyilmaz},
  \citenamefont {Castro~Neto}, \citenamefont {Xie}, \citenamefont {Fogler},\
  and\ \citenamefont {Basov}}]{RN19}%
  \BibitemOpen
  \bibfield  {author} {\bibinfo {author} {\bibfnamefont {G.}~\bibnamefont
  {Ni}}, \bibinfo {author} {\bibfnamefont {H.}~\bibnamefont {Wang}}, \bibinfo
  {author} {\bibfnamefont {J.}~\bibnamefont {Wu}}, \bibinfo {author}
  {\bibfnamefont {Z.}~\bibnamefont {Fei}}, \bibinfo {author} {\bibfnamefont
  {M.}~\bibnamefont {Goldflam}}, \bibinfo {author} {\bibfnamefont
  {F.}~\bibnamefont {Keilmann}}, \bibinfo {author} {\bibfnamefont
  {B.}~\bibnamefont {{\"O}zyilmaz}}, \bibinfo {author} {\bibfnamefont
  {A.}~\bibnamefont {Castro~Neto}}, \bibinfo {author} {\bibfnamefont
  {X.}~\bibnamefont {Xie}}, \bibinfo {author} {\bibfnamefont {M.}~\bibnamefont
  {Fogler}},\ and\ \bibinfo {author} {\bibfnamefont {D.}~\bibnamefont
  {Basov}},\ }\bibfield  {title} {\bibinfo {title} {Plasmons in graphene
  moir{\'e} superlattices},\ }\href {https://doi.org/10.1038/nmat4425}
  {\bibfield  {journal} {\bibinfo  {journal} {Nat. Mater.}\ }\textbf {\bibinfo
  {volume} {14}},\ \bibinfo {pages} {1217} (\bibinfo {year}
  {2015})}\BibitemShut {NoStop}%
\bibitem [{\citenamefont {Alnasser}\ \emph {et~al.}(2021)\citenamefont
  {Alnasser}, \citenamefont {Kamau}, \citenamefont {Hurley}, \citenamefont
  {Cui},\ and\ \citenamefont {Lin}}]{RN20}%
  \BibitemOpen
  \bibfield  {author} {\bibinfo {author} {\bibfnamefont {K.}~\bibnamefont
  {Alnasser}}, \bibinfo {author} {\bibfnamefont {S.}~\bibnamefont {Kamau}},
  \bibinfo {author} {\bibfnamefont {N.}~\bibnamefont {Hurley}}, \bibinfo
  {author} {\bibfnamefont {J.}~\bibnamefont {Cui}},\ and\ \bibinfo {author}
  {\bibfnamefont {Y.}~\bibnamefont {Lin}},\ }\bibfield  {title} {\bibinfo
  {title} {Photonic band gaps and resonance modes in 2d twisted moir{\'e}
  photonic crystal},\ }\href {https://doi.org/10.3390/photonics8100408}
  {\bibfield  {journal} {\bibinfo  {journal} {Photonics}\ }\textbf {\bibinfo
  {volume} {8}},\ \bibinfo {pages} {408} (\bibinfo {year} {2021})}\BibitemShut
  {NoStop}%
\bibitem [{\citenamefont {Wang}\ \emph
  {et~al.}(2019{\natexlab{a}})\citenamefont {Wang}, \citenamefont {Xie},
  \citenamefont {Gupta}, \citenamefont {Zhu}, \citenamefont {Liu},
  \citenamefont {Liu}, \citenamefont {Lu},\ and\ \citenamefont {Chen}}]{RN21}%
  \BibitemOpen
  \bibfield  {author} {\bibinfo {author} {\bibfnamefont {H.}~\bibnamefont
  {Wang}}, \bibinfo {author} {\bibfnamefont {B.}~\bibnamefont {Xie}}, \bibinfo
  {author} {\bibfnamefont {S.~K.}\ \bibnamefont {Gupta}}, \bibinfo {author}
  {\bibfnamefont {X.}~\bibnamefont {Zhu}}, \bibinfo {author} {\bibfnamefont
  {L.}~\bibnamefont {Liu}}, \bibinfo {author} {\bibfnamefont {X.}~\bibnamefont
  {Liu}}, \bibinfo {author} {\bibfnamefont {M.}~\bibnamefont {Lu}},\ and\
  \bibinfo {author} {\bibfnamefont {Y.}~\bibnamefont {Chen}},\ }\bibfield
  {title} {\bibinfo {title} {Exceptional concentric rings in a non-hermitian
  bilayer photonic system},\ }\href
  {https://doi.org/10.1103/PhysRevB.100.165134} {\bibfield  {journal} {\bibinfo
   {journal} {Phys. Rev. B}\ }\textbf {\bibinfo {volume} {100}},\ \bibinfo
  {pages} {165134} (\bibinfo {year} {2019}{\natexlab{a}})}\BibitemShut
  {NoStop}%
\bibitem [{\citenamefont {Yang}\ \emph {et~al.}(2023)\citenamefont {Yang},
  \citenamefont {Roques-Carmes}, \citenamefont {Kooi}, \citenamefont {Tang},
  \citenamefont {Beroz}, \citenamefont {Mazur}, \citenamefont {Kaminer},
  \citenamefont {Joannopoulos},\ and\ \citenamefont
  {Solja{\v{c}}i{\'c}}}]{RN22}%
  \BibitemOpen
  \bibfield  {author} {\bibinfo {author} {\bibfnamefont {Y.}~\bibnamefont
  {Yang}}, \bibinfo {author} {\bibfnamefont {C.}~\bibnamefont {Roques-Carmes}},
  \bibinfo {author} {\bibfnamefont {S.~E.}\ \bibnamefont {Kooi}}, \bibinfo
  {author} {\bibfnamefont {H.}~\bibnamefont {Tang}}, \bibinfo {author}
  {\bibfnamefont {J.}~\bibnamefont {Beroz}}, \bibinfo {author} {\bibfnamefont
  {E.}~\bibnamefont {Mazur}}, \bibinfo {author} {\bibfnamefont
  {I.}~\bibnamefont {Kaminer}}, \bibinfo {author} {\bibfnamefont {J.~D.}\
  \bibnamefont {Joannopoulos}},\ and\ \bibinfo {author} {\bibfnamefont
  {M.}~\bibnamefont {Solja{\v{c}}i{\'c}}},\ }\bibfield  {title} {\bibinfo
  {title} {Photonic flatband resonances for free-electron radiation},\ }\href
  {https://doi.org/10.1038/s41586-022-05387-5} {\bibfield  {journal} {\bibinfo
  {journal} {Nature}\ }\textbf {\bibinfo {volume} {613}},\ \bibinfo {pages}
  {42} (\bibinfo {year} {2023})}\BibitemShut {NoStop}%
\bibitem [{\citenamefont {Wu}\ \emph {et~al.}(2022)\citenamefont {Wu},
  \citenamefont {Lin}, \citenamefont {Jiang}, \citenamefont {Zhou},
  \citenamefont {Hang}, \citenamefont {Hou},\ and\ \citenamefont
  {Jiang}}]{RN23}%
  \BibitemOpen
  \bibfield  {author} {\bibinfo {author} {\bibfnamefont {S.}~\bibnamefont
  {Wu}}, \bibinfo {author} {\bibfnamefont {Z.}~\bibnamefont {Lin}}, \bibinfo
  {author} {\bibfnamefont {B.}~\bibnamefont {Jiang}}, \bibinfo {author}
  {\bibfnamefont {X.}~\bibnamefont {Zhou}}, \bibinfo {author} {\bibfnamefont
  {Z.~H.}\ \bibnamefont {Hang}}, \bibinfo {author} {\bibfnamefont
  {B.}~\bibnamefont {Hou}},\ and\ \bibinfo {author} {\bibfnamefont
  {J.}~\bibnamefont {Jiang}},\ }\bibfield  {title} {\bibinfo {title}
  {Higher-order topological states in acoustic twisted moir{\'e}
  superlattices},\ }\href {https://doi.org/10.1103/PhysRevApplied.17.034061}
  {\bibfield  {journal} {\bibinfo  {journal} {Phys. Rev. Appl.}\ }\textbf
  {\bibinfo {volume} {17}},\ \bibinfo {pages} {034061} (\bibinfo {year}
  {2022})}\BibitemShut {NoStop}%
\bibitem [{\citenamefont {Peng}\ \emph {et~al.}(2022)\citenamefont {Peng},
  \citenamefont {Liu}, \citenamefont {Yan}, \citenamefont {Peng}, \citenamefont
  {Shi}, \citenamefont {Xie}, \citenamefont {Li}, \citenamefont {Xiang},\ and\
  \citenamefont {Liu}}]{RN24}%
  \BibitemOpen
  \bibfield  {author} {\bibinfo {author} {\bibfnamefont {P.}~\bibnamefont
  {Peng}}, \bibinfo {author} {\bibfnamefont {E.}~\bibnamefont {Liu}}, \bibinfo
  {author} {\bibfnamefont {B.}~\bibnamefont {Yan}}, \bibinfo {author}
  {\bibfnamefont {Y.}~\bibnamefont {Peng}}, \bibinfo {author} {\bibfnamefont
  {A.}~\bibnamefont {Shi}}, \bibinfo {author} {\bibfnamefont {J.}~\bibnamefont
  {Xie}}, \bibinfo {author} {\bibfnamefont {H.}~\bibnamefont {Li}}, \bibinfo
  {author} {\bibfnamefont {Y.}~\bibnamefont {Xiang}},\ and\ \bibinfo {author}
  {\bibfnamefont {J.}~\bibnamefont {Liu}},\ }\bibfield  {title} {\bibinfo
  {title} {{Pair-partitioned bulk localized states induced by topological band
  inversion}},\ }\href {https://doi.org/10.1063/5.0101925} {\bibfield
  {journal} {\bibinfo  {journal} {Appl. Phys. Lett.}\ }\textbf {\bibinfo
  {volume} {121}},\ \bibinfo {pages} {011103} (\bibinfo {year}
  {2022})}\BibitemShut {NoStop}%
\bibitem [{\citenamefont {Yan}\ \emph {et~al.}(2024)\citenamefont {Yan},
  \citenamefont {Peng}, \citenamefont {Xie}, \citenamefont {Peng},
  \citenamefont {Shi}, \citenamefont {Li}, \citenamefont {Gao}, \citenamefont
  {Peng}, \citenamefont {Jiang}, \citenamefont {Liu}, \citenamefont {Gao},\
  and\ \citenamefont {Wen}}]{RN25}%
  \BibitemOpen
  \bibfield  {author} {\bibinfo {author} {\bibfnamefont {B.}~\bibnamefont
  {Yan}}, \bibinfo {author} {\bibfnamefont {Y.}~\bibnamefont {Peng}}, \bibinfo
  {author} {\bibfnamefont {J.}~\bibnamefont {Xie}}, \bibinfo {author}
  {\bibfnamefont {Y.}~\bibnamefont {Peng}}, \bibinfo {author} {\bibfnamefont
  {A.}~\bibnamefont {Shi}}, \bibinfo {author} {\bibfnamefont {H.}~\bibnamefont
  {Li}}, \bibinfo {author} {\bibfnamefont {F.}~\bibnamefont {Gao}}, \bibinfo
  {author} {\bibfnamefont {P.}~\bibnamefont {Peng}}, \bibinfo {author}
  {\bibfnamefont {J.}~\bibnamefont {Jiang}}, \bibinfo {author} {\bibfnamefont
  {J.}~\bibnamefont {Liu}}, \bibinfo {author} {\bibfnamefont {F.}~\bibnamefont
  {Gao}},\ and\ \bibinfo {author} {\bibfnamefont {S.}~\bibnamefont {Wen}},\
  }\bibfield  {title} {\bibinfo {title} {Multifrequency and multimode
  topological waveguides in a stampfli-triangle photonic crystal with large
  valley chern numbers},\ }\href
  {https://doi.org/https://doi.org/10.1002/lpor.202300686} {\bibfield
  {journal} {\bibinfo  {journal} {Laser Photonics Rev.}\ ,\ \bibinfo {pages}
  {2300686}} (\bibinfo {year} {2024})}\BibitemShut {NoStop}%
\bibitem [{\citenamefont {Shi}\ \emph {et~al.}(2024)\citenamefont {Shi},
  \citenamefont {Peng}, \citenamefont {Jiang}, \citenamefont {Peng},
  \citenamefont {Peng}, \citenamefont {Chen}, \citenamefont {Chen},
  \citenamefont {Wen}, \citenamefont {Lin}, \citenamefont {Gao},\ and\
  \citenamefont {Liu}}]{RN26}%
  \BibitemOpen
  \bibfield  {author} {\bibinfo {author} {\bibfnamefont {A.}~\bibnamefont
  {Shi}}, \bibinfo {author} {\bibfnamefont {Y.}~\bibnamefont {Peng}}, \bibinfo
  {author} {\bibfnamefont {J.}~\bibnamefont {Jiang}}, \bibinfo {author}
  {\bibfnamefont {Y.}~\bibnamefont {Peng}}, \bibinfo {author} {\bibfnamefont
  {P.}~\bibnamefont {Peng}}, \bibinfo {author} {\bibfnamefont {J.}~\bibnamefont
  {Chen}}, \bibinfo {author} {\bibfnamefont {H.}~\bibnamefont {Chen}}, \bibinfo
  {author} {\bibfnamefont {S.}~\bibnamefont {Wen}}, \bibinfo {author}
  {\bibfnamefont {X.}~\bibnamefont {Lin}}, \bibinfo {author} {\bibfnamefont
  {F.}~\bibnamefont {Gao}},\ and\ \bibinfo {author} {\bibfnamefont
  {J.}~\bibnamefont {Liu}},\ }\bibfield  {title} {\bibinfo {title} {Observation
  of topological corner state arrays in photonic quasicrystals},\ }\href
  {https://doi.org/https://doi.org/10.1002/lpor.202300956} {\bibfield
  {journal} {\bibinfo  {journal} {Laser Photonics Rev.}\ ,\ \bibinfo {pages}
  {2300956}} (\bibinfo {year} {2024})}\BibitemShut {NoStop}%
\bibitem [{\citenamefont {Xu}\ \emph {et~al.}(2024)\citenamefont {Xu},
  \citenamefont {Peng}, \citenamefont {Shi}, \citenamefont {Peng},\ and\
  \citenamefont {Liu}}]{RN27}%
  \BibitemOpen
  \bibfield  {author} {\bibinfo {author} {\bibfnamefont {Q.}~\bibnamefont
  {Xu}}, \bibinfo {author} {\bibfnamefont {Y.}~\bibnamefont {Peng}}, \bibinfo
  {author} {\bibfnamefont {A.}~\bibnamefont {Shi}}, \bibinfo {author}
  {\bibfnamefont {P.}~\bibnamefont {Peng}},\ and\ \bibinfo {author}
  {\bibfnamefont {J.}~\bibnamefont {Liu}},\ }\bibfield  {title} {\bibinfo
  {title} {Dual-band topological rainbows in penrose-triangle photonic
  crystals},\ }\href {https://doi.org/10.1364/JOSAA.507789} {\bibfield
  {journal} {\bibinfo  {journal} {J. Opt. Soc. Am. A}\ }\textbf {\bibinfo
  {volume} {41}},\ \bibinfo {pages} {366} (\bibinfo {year} {2024})}\BibitemShut
  {NoStop}%
\bibitem [{\citenamefont {Song}\ \emph {et~al.}(2020)\citenamefont {Song},
  \citenamefont {Sun},\ and\ \citenamefont {Wang}}]{RN28}%
  \BibitemOpen
  \bibfield  {author} {\bibinfo {author} {\bibfnamefont {Z.}~\bibnamefont
  {Song}}, \bibinfo {author} {\bibfnamefont {X.}~\bibnamefont {Sun}},\ and\
  \bibinfo {author} {\bibfnamefont {L.}~\bibnamefont {Wang}},\ }\bibfield
  {title} {\bibinfo {title} {Switchable asymmetric moir{\'e} patterns with
  strongly localized states},\ }\href
  {https://doi.org/10.1021/acs.jpclett.0c02400} {\bibfield  {journal} {\bibinfo
   {journal} {J. Phys. Chem. Lett.}\ }\textbf {\bibinfo {volume} {11}},\
  \bibinfo {pages} {9224} (\bibinfo {year} {2020})}\BibitemShut {NoStop}%
\bibitem [{\citenamefont {Wood}(1964)}]{RN29}%
  \BibitemOpen
  \bibfield  {author} {\bibinfo {author} {\bibfnamefont {E.~A.}\ \bibnamefont
  {Wood}},\ }\bibfield  {title} {\bibinfo {title} {Vocabulary of surface
  crystallography},\ }\href {https://doi.org/10.1063/1.1713610} {\bibfield
  {journal} {\bibinfo  {journal} {J. Appl. Phys.}\ }\textbf {\bibinfo {volume}
  {35}},\ \bibinfo {pages} {1306} (\bibinfo {year} {1964})}\BibitemShut
  {NoStop}%
\bibitem [{\citenamefont {Park}\ and\ \citenamefont {Stroud}(2004)}]{RN30}%
  \BibitemOpen
  \bibfield  {author} {\bibinfo {author} {\bibfnamefont {S.~Y.}\ \bibnamefont
  {Park}}\ and\ \bibinfo {author} {\bibfnamefont {D.}~\bibnamefont {Stroud}},\
  }\bibfield  {title} {\bibinfo {title} {Surface-plasmon dispersion relations
  in chains of metallic nanoparticles: An exact quasistatic calculation},\
  }\href {https://doi.org/10.1103/PhysRevB.69.125418} {\bibfield  {journal}
  {\bibinfo  {journal} {Phys. Rev. B}\ }\textbf {\bibinfo {volume} {69}},\
  \bibinfo {pages} {125418} (\bibinfo {year} {2004})}\BibitemShut {NoStop}%
\bibitem [{\citenamefont {Li}\ \emph {et~al.}(2020)\citenamefont {Li},
  \citenamefont {Zhirihin}, \citenamefont {Gorlach}, \citenamefont {Ni},
  \citenamefont {Filonov}, \citenamefont {Slobozhanyuk}, \citenamefont
  {Al{\`u}},\ and\ \citenamefont {Khanikaev}}]{RN31}%
  \BibitemOpen
  \bibfield  {author} {\bibinfo {author} {\bibfnamefont {M.}~\bibnamefont
  {Li}}, \bibinfo {author} {\bibfnamefont {D.}~\bibnamefont {Zhirihin}},
  \bibinfo {author} {\bibfnamefont {M.}~\bibnamefont {Gorlach}}, \bibinfo
  {author} {\bibfnamefont {X.}~\bibnamefont {Ni}}, \bibinfo {author}
  {\bibfnamefont {D.}~\bibnamefont {Filonov}}, \bibinfo {author} {\bibfnamefont
  {A.}~\bibnamefont {Slobozhanyuk}}, \bibinfo {author} {\bibfnamefont
  {A.}~\bibnamefont {Al{\`u}}},\ and\ \bibinfo {author} {\bibfnamefont {A.~B.}\
  \bibnamefont {Khanikaev}},\ }\bibfield  {title} {\bibinfo {title}
  {Higher-order topological states in photonic kagome crystals with long-range
  interactions},\ }\href {https://doi.org/10.1038/s41566-019-0561-9} {\bibfield
   {journal} {\bibinfo  {journal} {Nat. Photonics}\ }\textbf {\bibinfo {volume}
  {14}},\ \bibinfo {pages} {89} (\bibinfo {year} {2020})}\BibitemShut {NoStop}%
\bibitem [{\citenamefont {Bouhon}\ \emph {et~al.}(2019)\citenamefont {Bouhon},
  \citenamefont {Black-Schaffer},\ and\ \citenamefont {Slager}}]{RN32}%
  \BibitemOpen
  \bibfield  {author} {\bibinfo {author} {\bibfnamefont {A.}~\bibnamefont
  {Bouhon}}, \bibinfo {author} {\bibfnamefont {A.~M.}\ \bibnamefont
  {Black-Schaffer}},\ and\ \bibinfo {author} {\bibfnamefont {R.-J.}\
  \bibnamefont {Slager}},\ }\bibfield  {title} {\bibinfo {title} {Wilson loop
  approach to fragile topology of split elementary band representations and
  topological crystalline insulators with time-reversal symmetry},\ }\href
  {https://doi.org/10.1103/PhysRevB.100.195135} {\bibfield  {journal} {\bibinfo
   {journal} {Phys. Rev. B}\ }\textbf {\bibinfo {volume} {100}},\ \bibinfo
  {pages} {195135} (\bibinfo {year} {2019})}\BibitemShut {NoStop}%
\bibitem [{\citenamefont {Wang}\ \emph
  {et~al.}(2019{\natexlab{b}})\citenamefont {Wang}, \citenamefont {Guo},\ and\
  \citenamefont {Jiang}}]{RN33}%
  \BibitemOpen
  \bibfield  {author} {\bibinfo {author} {\bibfnamefont {H.-X.}\ \bibnamefont
  {Wang}}, \bibinfo {author} {\bibfnamefont {G.-Y.}\ \bibnamefont {Guo}},\ and\
  \bibinfo {author} {\bibfnamefont {J.-H.}\ \bibnamefont {Jiang}},\ }\bibfield
  {title} {\bibinfo {title} {Band topology in classical waves: Wilson-loop
  approach to topological numbers and fragile topology},\ }\href
  {https://doi.org/10.1088/1367-2630/ab3f71} {\bibfield  {journal} {\bibinfo
  {journal} {New J. Phys.}\ }\textbf {\bibinfo {volume} {21}},\ \bibinfo
  {pages} {093029} (\bibinfo {year} {2019}{\natexlab{b}})}\BibitemShut
  {NoStop}%
\bibitem [{\citenamefont {Zhang}\ \emph
  {et~al.}(2022{\natexlab{b}})\citenamefont {Zhang}, \citenamefont {Tang},
  \citenamefont {Dai}, \citenamefont {Zhang}, \citenamefont {Cao},\ and\
  \citenamefont {Xiang}}]{RN34}%
  \BibitemOpen
  \bibfield  {author} {\bibinfo {author} {\bibfnamefont {Y.}~\bibnamefont
  {Zhang}}, \bibinfo {author} {\bibfnamefont {J.}~\bibnamefont {Tang}},
  \bibinfo {author} {\bibfnamefont {X.}~\bibnamefont {Dai}}, \bibinfo {author}
  {\bibfnamefont {S.}~\bibnamefont {Zhang}}, \bibinfo {author} {\bibfnamefont
  {Z.}~\bibnamefont {Cao}},\ and\ \bibinfo {author} {\bibfnamefont
  {Y.}~\bibnamefont {Xiang}},\ }\bibfield  {title} {\bibinfo {title} {Design of
  a higher-order nodal-line semimetal in a spring-shaped acoustic topological
  crystal},\ }\href {https://doi.org/10.1103/PhysRevB.106.184101} {\bibfield
  {journal} {\bibinfo  {journal} {Phys. Rev. B}\ }\textbf {\bibinfo {volume}
  {106}},\ \bibinfo {pages} {184101} (\bibinfo {year}
  {2022}{\natexlab{b}})}\BibitemShut {NoStop}%
\bibitem [{\citenamefont {Zhang}\ \emph {et~al.}(2023)\citenamefont {Zhang},
  \citenamefont {Tang}, \citenamefont {Dai},\ and\ \citenamefont
  {Xiang}}]{RN35}%
  \BibitemOpen
  \bibfield  {author} {\bibinfo {author} {\bibfnamefont {Y.}~\bibnamefont
  {Zhang}}, \bibinfo {author} {\bibfnamefont {J.}~\bibnamefont {Tang}},
  \bibinfo {author} {\bibfnamefont {X.}~\bibnamefont {Dai}},\ and\ \bibinfo
  {author} {\bibfnamefont {Y.}~\bibnamefont {Xiang}},\ }\bibfield  {title}
  {\bibinfo {title} {Flexible dimensional hierarchy of higher-order topology in
  the stacked kagome-chain acoustic crystal},\ }\href
  {https://doi.org/10.1038/s42005-023-01254-5} {\bibfield  {journal} {\bibinfo
  {journal} {Commun. Phys.}\ }\textbf {\bibinfo {volume} {6}},\ \bibinfo
  {pages} {130} (\bibinfo {year} {2023})}\BibitemShut {NoStop}%
\bibitem [{\citenamefont {Elcoro}\ \emph {et~al.}(2017)\citenamefont {Elcoro},
  \citenamefont {Bradlyn}, \citenamefont {Wang}, \citenamefont {Vergniory},
  \citenamefont {Cano}, \citenamefont {Felser}, \citenamefont {Bernevig},
  \citenamefont {Orobengoa}, \citenamefont {Flor},\ and\ \citenamefont
  {Aroyo}}]{RN36}%
  \BibitemOpen
  \bibfield  {author} {\bibinfo {author} {\bibfnamefont {L.}~\bibnamefont
  {Elcoro}}, \bibinfo {author} {\bibfnamefont {B.}~\bibnamefont {Bradlyn}},
  \bibinfo {author} {\bibfnamefont {Z.}~\bibnamefont {Wang}}, \bibinfo {author}
  {\bibfnamefont {M.~G.}\ \bibnamefont {Vergniory}}, \bibinfo {author}
  {\bibfnamefont {J.}~\bibnamefont {Cano}}, \bibinfo {author} {\bibfnamefont
  {C.}~\bibnamefont {Felser}}, \bibinfo {author} {\bibfnamefont {B.~A.}\
  \bibnamefont {Bernevig}}, \bibinfo {author} {\bibfnamefont {D.}~\bibnamefont
  {Orobengoa}}, \bibinfo {author} {\bibfnamefont {G.}~\bibnamefont {Flor}},\
  and\ \bibinfo {author} {\bibfnamefont {M.~I.}\ \bibnamefont {Aroyo}},\
  }\bibfield  {title} {\bibinfo {title} {Double crystallographic groups and
  their representations on the bilbao crystallographic server},\ }\href
  {https://doi.org/10.1107/S1600576717011712} {\bibfield  {journal} {\bibinfo
  {journal} {J. Appl. Cryst.}\ }\textbf {\bibinfo {volume} {50}},\ \bibinfo
  {pages} {1457} (\bibinfo {year} {2017})}\BibitemShut {NoStop}%
\bibitem [{\citenamefont {Slager}\ \emph {et~al.}(2013)\citenamefont {Slager},
  \citenamefont {Mesaros}, \citenamefont {Juricic},\ and\ \citenamefont
  {Zaanen}}]{RN37}%
  \BibitemOpen
  \bibfield  {author} {\bibinfo {author} {\bibfnamefont {R.-J.}\ \bibnamefont
  {Slager}}, \bibinfo {author} {\bibfnamefont {A.}~\bibnamefont {Mesaros}},
  \bibinfo {author} {\bibfnamefont {V.}~\bibnamefont {Juricic}},\ and\ \bibinfo
  {author} {\bibfnamefont {J.}~\bibnamefont {Zaanen}},\ }\bibfield  {title}
  {\bibinfo {title} {The space group classification of topological
  band-insulators},\ }\href {https://doi.org/10.1038/nphys2513} {\bibfield
  {journal} {\bibinfo  {journal} {Nat. Phys.}\ }\textbf {\bibinfo {volume}
  {9}},\ \bibinfo {pages} {98} (\bibinfo {year} {2013})}\BibitemShut {NoStop}%
\bibitem [{\citenamefont {Kruthoff}\ \emph {et~al.}(2017)\citenamefont
  {Kruthoff}, \citenamefont {de~Boer}, \citenamefont {van Wezel}, \citenamefont
  {Kane},\ and\ \citenamefont {Slager}}]{RN38}%
  \BibitemOpen
  \bibfield  {author} {\bibinfo {author} {\bibfnamefont {J.}~\bibnamefont
  {Kruthoff}}, \bibinfo {author} {\bibfnamefont {J.}~\bibnamefont {de~Boer}},
  \bibinfo {author} {\bibfnamefont {J.}~\bibnamefont {van Wezel}}, \bibinfo
  {author} {\bibfnamefont {C.~L.}\ \bibnamefont {Kane}},\ and\ \bibinfo
  {author} {\bibfnamefont {R.-J.}\ \bibnamefont {Slager}},\ }\bibfield  {title}
  {\bibinfo {title} {Topological classification of crystalline insulators
  through band structure combinatorics},\ }\href
  {https://doi.org/10.1103/PhysRevX.7.041069} {\bibfield  {journal} {\bibinfo
  {journal} {Phys. Rev. X}\ }\textbf {\bibinfo {volume} {7}},\ \bibinfo {pages}
  {041069} (\bibinfo {year} {2017})}\BibitemShut {NoStop}%
\bibitem [{\citenamefont {Bradlyn}\ \emph {et~al.}(2017)\citenamefont
  {Bradlyn}, \citenamefont {Elcoro}, \citenamefont {Cano}, \citenamefont
  {Vergniory}, \citenamefont {Wang}, \citenamefont {Felser}, \citenamefont
  {Aroyo},\ and\ \citenamefont {Bernevig}}]{RN39}%
  \BibitemOpen
  \bibfield  {author} {\bibinfo {author} {\bibfnamefont {B.}~\bibnamefont
  {Bradlyn}}, \bibinfo {author} {\bibfnamefont {L.}~\bibnamefont {Elcoro}},
  \bibinfo {author} {\bibfnamefont {J.}~\bibnamefont {Cano}}, \bibinfo {author}
  {\bibfnamefont {M.~G.}\ \bibnamefont {Vergniory}}, \bibinfo {author}
  {\bibfnamefont {Z.}~\bibnamefont {Wang}}, \bibinfo {author} {\bibfnamefont
  {C.}~\bibnamefont {Felser}}, \bibinfo {author} {\bibfnamefont {M.~I.}\
  \bibnamefont {Aroyo}},\ and\ \bibinfo {author} {\bibfnamefont {B.~A.}\
  \bibnamefont {Bernevig}},\ }\bibfield  {title} {\bibinfo {title} {Topological
  quantum chemistry},\ }\href {https://doi.org/10.1038/nature23268} {\bibfield
  {journal} {\bibinfo  {journal} {Nature}\ }\textbf {\bibinfo {volume} {547}},\
  \bibinfo {pages} {298} (\bibinfo {year} {2017})}\BibitemShut {NoStop}%
\bibitem [{\citenamefont {Benalcazar}\ \emph {et~al.}(2019)\citenamefont
  {Benalcazar}, \citenamefont {Li},\ and\ \citenamefont {Hughes}}]{RN40}%
  \BibitemOpen
  \bibfield  {author} {\bibinfo {author} {\bibfnamefont {W.~A.}\ \bibnamefont
  {Benalcazar}}, \bibinfo {author} {\bibfnamefont {T.}~\bibnamefont {Li}},\
  and\ \bibinfo {author} {\bibfnamefont {T.~L.}\ \bibnamefont {Hughes}},\
  }\bibfield  {title} {\bibinfo {title} {Quantization of fractional corner
  charge in ${C}_{n}$-symmetric higher-order topological crystalline
  insulators},\ }\href {https://doi.org/10.1103/PhysRevB.99.245151} {\bibfield
  {journal} {\bibinfo  {journal} {Phys. Rev. B}\ }\textbf {\bibinfo {volume}
  {99}},\ \bibinfo {pages} {245151} (\bibinfo {year} {2019})}\BibitemShut
  {NoStop}%
\bibitem [{\citenamefont {S.~Vaidya}\ and\ \citenamefont
  {Benalcazar}(2023)}]{RN41}%
  \BibitemOpen
  \bibfield  {author} {\bibinfo {author} {\bibfnamefont {T.~C. M. C.~R.}\
  \bibnamefont {S.~Vaidya}, \bibfnamefont {A.~Ghorashi}}\ and\ \bibinfo
  {author} {\bibfnamefont {W.~A.}\ \bibnamefont {Benalcazar}},\ }\bibfield
  {title} {\bibinfo {title} {Topological phases of photonic crystals under
  crystalline symmetries},\ }\href
  {https://doi.org/10.1103/PhysRevB.108.085116} {\bibfield  {journal} {\bibinfo
   {journal} {Phys. Rev. B}\ }\textbf {\bibinfo {volume} {108}},\ \bibinfo
  {pages} {085116} (\bibinfo {year} {2023})}\BibitemShut {NoStop}%
\end{thebibliography}%
 	%

\end{document}